\documentclass[aps,pra,notitlepage,superscriptaddress,longbibliography,10pt]{revtex4-1}
\setcitestyle{super}
\usepackage[colorlinks=true, allcolors=blue]{hyperref}
\usepackage{amssymb}

\usepackage{etoolbox}		
	\patchcmd{\section}{\centering}{\hspace*{3.75em}\raggedright}{}{}
	\patchcmd{\subsection}{\centering}{\hspace*{2em}\raggedright}{}{}
	\patchcmd{\subsubsection}{\centering}{\hspace*{0.25em}\raggedright}{}{}
	\patchcmd{\subsubsection}{\normalfont\small\itshape}{\normalfont\small\itshape\bfseries}{}{}
\makeatletter				
	\renewcommand*{\thesection}{\arabic{section}}
	
	\renewcommand*{\p@subsection}{}
	
	\renewcommand*{\p@subsubsection}{}
\makeatother

\usepackage{amsmath,bm,amssymb,esint}			
\usepackage{lmodern}							
\usepackage{enumitem}
\usepackage{graphicx}
\graphicspath{{./Figures/}}

\usepackage{xcolor}

\usepackage{lineno}

\newcommand{\be}{\begin{equation}}
\newcommand{\ee}{\end{equation}}
\newcommand{\BE}{\begin{eqnarray}}
\newcommand{\EE}{\end{eqnarray}}

\newcommand{\Bd}{\textbf{Bd}}
\newcommand{\bD}{\textbf{bD}}
\newcommand{\dB}{\textbf{dB}}
\newcommand{\Db}{\textbf{Db}}
\newcommand{\BD}{\textbf{BD}}
\newcommand{\DB}{\textbf{DB}}

\newcommand{\TITLE}
{Motion, fixation probability and the choice of an evolutionary process}

\begin{document}
\title{\TITLE}

\author{Francisco Herrer\'{i}as-Azcu\'{e}}
	\email{francisco.herreriasazcue@manchester.ac.uk}
	\affiliation{Theoretical Physics, School of Physics and Astronomy, The University of Manchester, Manchester M13 9PL, United Kingdom}
\author{Vicente P\'{e}rez-Mu\~{n}uzuri}
	\email{vicente.perez@cesga.es}
	\affiliation{Group of Nonlinear Physics, Faculty of Physics, University of Santiago de Compostela E-15782 Santiago de Compostela, Spain}
\author{Tobias Galla}
	\email{tobias.galla@manchester.ac.uk}
	\affiliation{Theoretical Physics, School of Physics and Astronomy, The University of Manchester, Manchester M13 9PL, United Kingdom}


		\begin{abstract}

Different evolutionary models are known to make disparate predictions for the success of an invading mutant in some situations. For example, some evolutionary mechanics lead to amplification of selection in structured populations, while others suppress it. Here, we use computer simulations to study evolutionary populations moved by flows, and show how the speed of this motion impacts the fixation probability of an invading mutant. Flows of different speeds interpolate between evolutionary dynamics on fixed heterogeneous graphs and in well-stirred populations. We find that the motion has an active role in amplifying or suppressing selection, accomplished by fragmenting and reconnecting the interaction graph. While increasing flow speeds suppress selection for most evolutionary models, we identify characteristic responses to flow for the different update rules we test. We suggest these responses as a potential aid for choosing the most suitable update rule for a given biological system. 

		\end{abstract}


\maketitle


\section{Introduction}\label{sec:Intro}
Over the years, different models have been proposed to describe evolutionary dynamics; they all describe how populations composed of members of different species change with time. These changes occur through death and birth events, and typically involve competition between individuals\cite{MaynardSmith1982, Ewens2004, Nowak2006a}. The models differ in the order in which these events are implemented and how competition takes place. For example, one distinguishes between `birth-death' and `death-birth' processes, or between global and local selection \cite{Nowak2006a,Zukewich2013,Kaveh2015}. Even though these seem to be relatively minor details, under certain circumstances they can lead to rather disparate outcomes\cite{Zukewich2013, Kaveh2015, Hindersin2015}. Choosing the right evolutionary model for a given biological system is therefore of great importance, but it is not a simple task.

There appears to be some tendency in the literature to simply choose one model and not argue in detail whether, for example, a birth-death or a death-birth process is better suited for the particular application at hand. While one could argue that most of these models are so stylised that the subtle differences between them are unimportant, significant differences in their predicted outcomes make it desirable to be able to distinguish between the options. This would enable one to make a more informed choice, and to identify the model which best captures the known behaviour of a given biological system.

In this paper we propose that such a possibility may exist if the members of the population are advected by an external flow. Specifically, we focus on the stochastic dynamics of a population initially composed only of wildtype individuals and a single invading mutant. The mutant will either be eliminated or its offspring will take over the population. We are interested in the latter outcome, and study the rate of successful fixation. We show that the response of the fixation probability to the flow speed can be very different for each evolutionary process. We speculate that this may be used to discriminate between different stylised models. In some experimental settings flow can be controlled externally, or situations without flow can be compared to those with fast flows. Differences in fixation probabilities have been found in static versus stirred populations of {\em Escherichia coli} for example\cite{Kerr2002,Habets2006,Perfeito2008a}. If such data is available, systematically studying the behaviour of different computational models of evolution in flowing populations can help to select the update mechanism which best captures the features of the biological system at hand.

\bigskip
The simplest approach to modelling stochastic evolution dispenses entirely with the notion of space and population structure, and assumes that the only relevant factors are the frequencies with which the different types of individuals are found in the population \cite{MaynardSmith1982,Ewens2004,Nowak2006a}. Each individual in such an unstructured population can interact with all other individuals at all times. To further simplify matters, birth and death events are usually coupled, so that one is immediately followed by the other; this facilitates the mathematical analysis, as it keeps the size of the population constant\cite{MaynardSmith1982, Ewens2004, Nowak2006a}.

If individuals are distributed in space, and have a limited range of interaction, the population becomes structured. Not every individual can interact with every other individual at all times. It is then helpful to consider the interaction graph of the population \cite{Nowak1992, Ebel2002, Nowak2006, Santos2006, Ohtsuki2006, Nowak2010, Szabo2007, Gross2008, Poncela2009, Masuda2009a, Broom2011, Perc2012, Wang2013, Perc2013, Hindersin2014, Jiang2014, Allen2017, Pavlogiannisa2018, Lieberman2005, Shakarian2012}. Nodes of these networks represent individuals, and links between nodes stand for potential interactions. Birth and death events take place between neighbouring nodes, that is, pairs of individuals connected by a link. The case of an unstructured population is recovered if links exist between any two individuals at all times; the interaction graph is then said to be complete.

Population structure has the potential to change the dynamics of evolutionary processes \cite{Nowak1992, Ebel2002, Nowak2006, Santos2006, Ohtsuki2006, Nowak2010, Szabo2007, Gross2008, Poncela2009, Masuda2009a, Broom2011, Perc2012, Wang2013, Perc2013, Hindersin2014, Jiang2014, Allen2017, Pavlogiannisa2018, Lieberman2005, Shakarian2012}. For example, species that would be selected against in an unstructured population are found to organise in clusters on networks, and in this way they can coexist with fitter types, or even eradicate the resident species. The time it takes for a species to reach fixation can be reduced or increased on networks. Evolution on simple graphs has been characterised mathematically (see for instance\cite{Lieberman2005,Broom2008,Broom2009,Houchmandzadeh2011a,Shakarian2012,Pattni2015} and references therein), but on more complicated networks the dynamics become much harder to describe analytically.

Further complication arises if the members of the populations are in motion. The interaction graph then becomes dynamic, making mathematical approaches more difficult. At the same time motion is a ubiquitous feature of biological systems, for example due to self-propulsion of microswimmers by means of flagella \cite{Lushi2014}, or advection of bacteria in a fluid environment \cite{Lapin2006}.

Common ways of implementing motion in computer models include migration; in such models individuals move to neighbouring sites on the interaction graph \cite{Nagylaki1980, McPeek1992, Hill2002a, Casagrandi2006, Houchmandzadeh2011, Thalhauser2015, Krieger2017}. Alternatively, adaptive networks have been considered; in these the link connecting two individuals is re-wired to a different individual, usually with preference for links between individuals of similar types \cite{Zimmermann2004, Zimmermann2005, Gross2006, Ehrhardt2006, Pacheco2006}. For further examples see also ref.~\cite{Gross2008} and references therein.
In this work, instead, we focus on populations that are not self-propelled, and use the type of motion one could expect in dynamic gaseous or aqueous environments. Specifically, the motion is due to a flow of the medium in which the population resides. The movement is not constrained by the current interaction network, and the interaction graph itself is dynamic. Evolutionary systems with this type of motion have been studied comparatively little; existing work includes refs.\cite{Krieger2017, Karolyi2005, Perlekar2011, Pigolotti2012, Pigolotti2012a, Pigolotti2013, Pigolotti2014, Plummer2018, Minors2018}. In ref.~\cite{Herrerias2018} we investigated stirred populations, and presented analytical solutions for the limit of very fast flows. Naively, one would expect that the success of a mutant under fast stirring is the same as the one on a complete graph, a situation also referred to as `well mixed' \cite{Ohtsuki2006, Nowak2010, Santos2006}. However, the results of ref.~\cite{Herrerias2018} showed that the fixation probability of an invading mutant approaches a different limiting value for very fast flows.

In this paper we investigate in more detail the effect of the flow speed on the fixation probability of invading mutants. In particular we also focus on intermediate and slow flows. We find that the way in which the flow affects the success of mutants depends on the choice of the evolutionary update rules.
We identify three main contributing factors: how well connected the initial mutant is with the rest of the population, the opportunities mutants have to organise in clustered groups, and how long individuals remain connected for as the flow moves them in space. These factors are influenced by the speed of the flow and, depending on the evolutionary update rule, they can either amplify or suppress selection relative to unstructured populations. This suggests the response of the fixation probability to flow speed as an indicator of the underlying evolutionary process. We think this can be a useful aid for choosing the most appropriate evolutionary model for given biological applications.

\section{Methods}\label{sec:Methods}
We use the same setup as ref.\cite{Herrerias2018}, and consider a population of fixed size $N$ composed of individuals of two species (wildtype and mutant). Unless specified otherwise, we use $N=100$. Individuals take positions in space within the two-dimensional domain $0\le x,y<1$ with periodic boundary conditions. Particles are subject to a continuous-time flow, moving them around in space, and to evolutionary dynamics, which change the frequencies of the two species in the population.

The motion of the particles is simulated through the so-called parallel shear flow\cite{Ottino1989,Neufeld2010}; we discuss the validity of our results for different flow fields in Sec.~\ref{sec:Discussion}. The velocity field of this flow is periodic in time, except for a random phase described below. During the first half of each period particles are moved vertically; the speed of each individual depends on the horizontal component of their position. During the second half of the period individuals move horizontally, with speeds dependent on their vertical positions. We write $v_x(x,y,t)$ and $v_y(x,y,t)$ for the velocity components of a particle at position $(x,y)$ at time $t$. Specifically, we use
\begin{align*}
	v_y(x,y,t)=0,~&~v_x(x,y,t)= V_{\rm max}\sin\left[2\pi y+\psi\right], && \hspace{-6em} \mbox{for~} t\in[nT,nT+T/2),\\
	v_x(x,y,t)=0,~&~v_y(x,y,t)= V_{\rm max}\sin\left[2\pi x+\psi\right], && \hspace{-6em} \mbox{for~} t\in[nT+T/2,(n+1)T),
\end{align*}
with $n=0,1,2,...$. The constant $V_{\rm max}$ sets the amplitude of the flow, and $T$ the period. The phase $\psi$ is drawn randomly from the interval $[0,2\pi)$ at the beginning of each half-period. Due to this random phase, the flow mimics chaotic motion; the trajectories of individuals who are initially close to each other diverge over time. At long times, the distribution of individuals moved by this flow is uniform in space \cite{Ottino1989,Neufeld2010}.

The evolutionary process is implemented through coupled birth and death events. The order in which reproduction and removal take place is important, and so we will distinguish between birth-death and death-birth processes. The evolutionary dynamics occur on an undirected interaction graph, dynamically generated by the flow. Specifically, we will say that one individual is a neighbour of another if they are within a distance $R$ of each other. 

\medskip
Individuals are in continuous motion, but evolutionary events occur at discrete times, ${t=\Delta t, ~2\Delta t,...}$ in our model. Simulations are then implemented as follows:
\begin{enumerate}[nolistsep]
\item At $t=0$, a population of $N$ particles is placed into the spatial domain at designated initial positions. These define an initial interaction graph. Of these individuals, $N-1$ are wildtype and one is a mutant. The mutant is chosen uniformly at random from the population.
\item The individuals are moved by the flow for a time interval $\Delta t$, leading to a new interaction graph.
\item An individual is chosen from the entire population. In the case of a birth-death process, it is designated to reproduce; for death-birth processes it is designated to die.
\item One of the neighbours of this individual is chosen to be replaced (birth-death) or to reproduce (death-birth).
\item The individual chosen for death adopts the species (wildtype or mutant) of the reproducing individual.
\item Repeat from step 2.
\end{enumerate}

\medskip
In each evolutionary step two individuals are chosen. They can either be \textit{picked} at random or \textit{selected} proportional to fitness. For the latter case we focus on frequency-independent selection; we set the wildtype fitness to one, and write $r$ for the fitness of the mutant species. Consider for example a group of $n_w$ wildtype individuals and $n_m$ mutants. A mutant would be selected to reproduce from this group with probability $rn_m/(rn_m+n_w)$, or a wildtype with probability $n_w/(rn_m+n_w)$. If selection is for death we proceed similarly, but with $r$ replaced by $1/r$. In this way, mutants are less likely to die than wildypes if $r>1$. For $r<1$ the mutant species is selected against. The simulation results shown in this paper focus on advantageous mutants; we set $r=1.05$ throughout.

Selection proportional to fitness can take place either in step 3 of the above algorithm (when an individual is chosen from the entire population) or in step 4 (when it is chosen from the neighbours of an individual). We refer to these cases as \textit{global} and \textit{local} selection, respectively. Since we distinguish between birth-death and death-birth processes, four combinations are possible: global birth-death (\Bd), global death-birth (\Db), local birth-death (\bD) and local death-birth (\dB). The capital letter in these acronyms indicates that selection dependent on fitness occurs in the respective step. 
In principle, one could also consider processes in which individuals are chosen proportional to fitness in both steps of the algorithm (\BD, \DB)\cite{Zukewich2013,Kaveh2015}. In order to be able to disentangle the effects that the flow has on fixation probabilities due to local or global selection, we limit the discussion in the main text to scenarios in which selection acts either globally or locally, but not both. The \BD ~and \DB ~processes are discussed in the Supporting Information.

\begin{figure}[t]
	\center
	\includegraphics[width=0.65\textwidth]{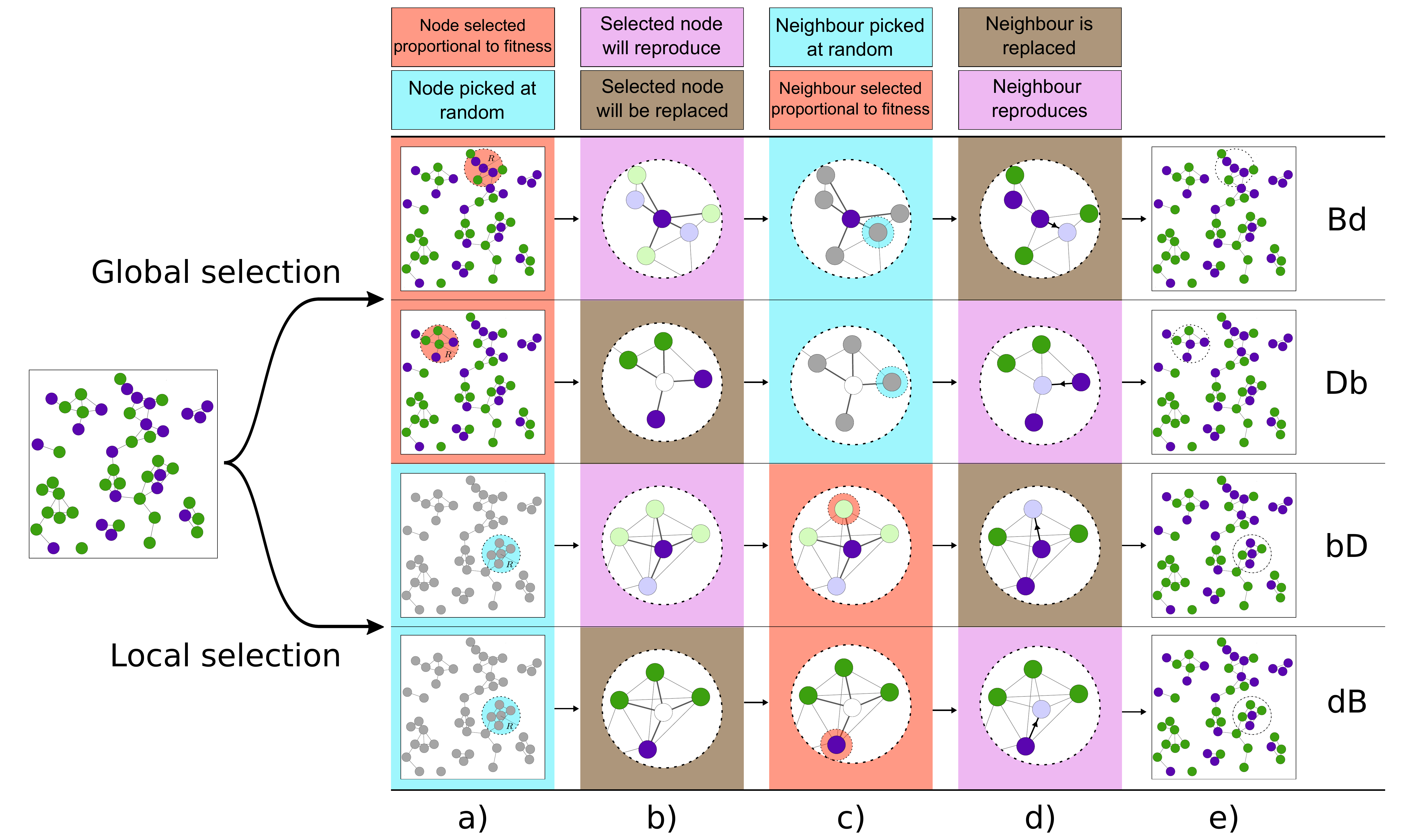}
	\caption{\textbf{Illustration of the update rules.} 
	Each row represents one of the different evolutionary update mechanisms. The columns indicate the different steps of each evolutionary event. 
	In column a) an individual is chosen from the whole population; it can be \textit{`selected'} through competition by fitness (red shading), or \textit{`picked'} at random, irrespective of its species (blue shading). 
	This node is destined to either reproduce (pink shading), or to be replaced (brown shading), as shown in column b).  
	Column c) indicates that one neighbour of this node is either \textit{selected} (red), or \textit{picked} (blue).
	This second node is destined to reproduce (pink), or to be replaced (brown), shown in column d).
	Column e) shows the result of the evolutionary event; the node chosen to reproduce places an offspring in place of the node chosen to die. Each row is composed of one box of each colour; the sequence of the colours distinguishes the different processes. 
	From top to bottom, the rows correspond to:
	(i) global birth-death process (\Bd): an individual is \textit{selected} from the whole population to reproduce, and one of its neighbours is \textit{picked} to be replaced by the first individual's offspring;
	(ii) global death-birth process (\Db): an individual is \textit{selected} to die from the whole population, and one of its neighbours is \textit{picked} to place an offspring in its place;
	(iii) local birth-death process (\bD): an individual is \textit{picked} from the whole population to reproduce, and one of its neighbours is \textit{selected} to die;
	(iv) local death-birth process (\dB): an individual is \textit{picked} from the whole population to die, and one of its neighbours is \textit{selected} to reproduce.
			}
	\label{fig:Processes}
\end{figure}

We illustrate the different evolutionary processes in Fig.~\ref{fig:Processes}. The upper two rows correspond to processes in which competition takes place among the entire population (\textit{global selection}). In the lower two rows the first node is picked irrespective of fitness, and competition takes place only among the neighbours of this node (\textit{local selection}). A step-by-step description of each of the processes can be found in the figure caption.

\medskip
One important characteristic of the flow is the typical timescale over which the set of neighbours of a given individual is renewed. More precisely, the set of neighbours of a given individual at time $t$, and at a later time $t+\tau$, will be uncorrelated provided $\tau$ is sufficiently large (see ref.\cite{Herrerias2018}, and the Supporting Information). This renewal time is in turn determined by the parameters $V_{\rm max}$, $T$ and $R$; following refs.\cite{Perez-Munuzuri2007,Galla2016}, we use $V_{\rm max}=1.4$ and $T=1$ throughout, and choose an interaction radius of $R=0.11$.

This choice of parameters leads to an estimate for the network renewal time of $\tau\approx 6.4$ (see Sec.~\ref{app:Timescales} in the Supporting Information for details). That is, the set of neighbours of one individual at one time is uncorrelated from its set of neighbours approximately six and a half flow periods earlier. It remains to specify how frequent evolutionary events are, i.e. to define the time step $\Delta t$ in the simulation described above. We treat this as a model parameter, and use $S=N \Delta t/T$ to quantify the number of generations elapsed in one flow period. Thus, $S$ indicates the speed of the flow relative to that of evolution. For small $S$, individuals move relatively little between evolutionary events (`slow flow'). Large values of $S$ describe fast flows.  From here on, we will refer to $S$ as the \textit{speed} of the flow, and investigate the outcome of evolution for different choices of this parameter. The flow speed $S$ is understood throughout as relative to the rate of evolutionary events. We note that the inverse of $S$ is related to the Damk\"{o}hler number in fluid dynamics \cite{Young2001,Sandulescu2007,Neufeld2002}.

\section{Results}\label{sec:Results}

\subsection{Effects of the flow speed on the fixation probability}\label{sec:Random}
We first address the case in which the initial coordinates of each individual are drawn from a uniform distribution on the domain $0\leq x,y<1$. The initial interaction graph is then a random geometric graph (RGG)\cite{Gilbert1961}.

For any non-zero flow rate ($S>0$) any member of the population can eventually interact with any other individual, even if they were not connected on the initial interaction graph. This is due to the mixing properties of the flow, and means that no individual can indefinitely remain isolated from the rest of the population. As a consequence, the final outcome of the evolutionary process is either fixation or extinction of the mutant.

\begin{figure}[t]
	\center
	\includegraphics[width=0.65\textwidth]{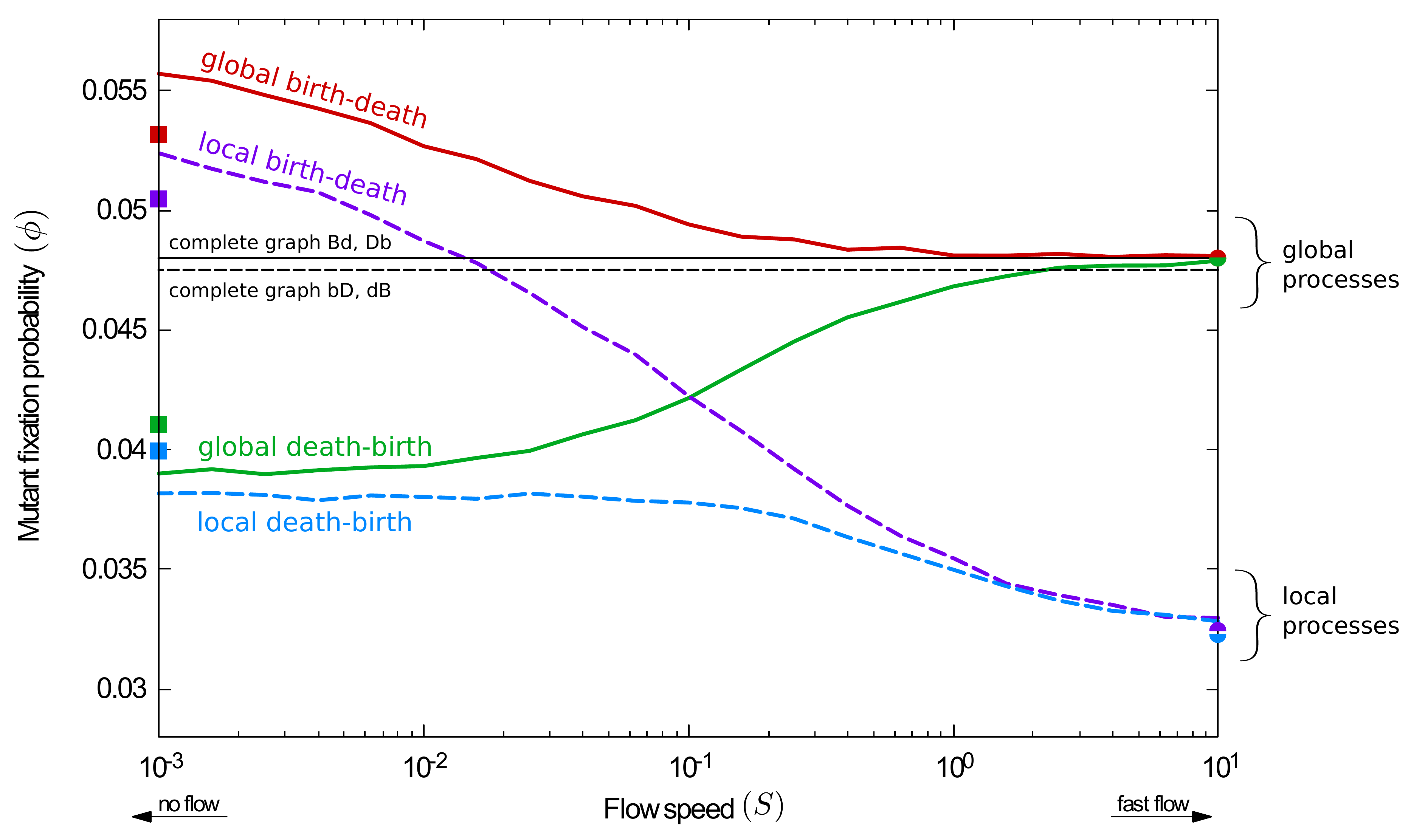}
	\caption{\textbf{Fixation probability as a function of the flow speed for unrestricted random initial positions (random geometric graphs, RGGs).} 
 For the global death-birth process, increasing the flow speed increases the fixation probability. The reverse is found for the remaining three processes. Circle markers show fixation probabilities in the fast-flow limit; square markers are results for fixed connected random geometric graphs (CRGGs); see text for further details. The fixation probabilities on a complete graph are shown for reference.
	}
	\label{fig:FixProbS_R}
\end{figure}

The fixation probability, $\phi$, for a beneficial mutation is depicted in Fig.~\ref{fig:FixProbS_R} as a function of the flow speed,~$S$. We show simulation results for the four different evolutionary processes \bD, \dB, \Bd, and \Db. Each data point is obtained from an ensemble of realisations. For comparison, we also show the fixation probability on a complete graph,~$\phi_{\rm {CG}}$. By definition, $\phi_{\rm CG}$ is independent of the flow speed, as all individuals interact with all others at all times. On complete graphs the fixation probability for global and local selection processes differ by a small amount\cite{Kaveh2015}.

Several interesting features can be observed in Fig.~\ref{fig:FixProbS_R}: 
For slow flows, the order of reproduction and removal is found to have a strong effect on the fixation probability, and it is less relevant whether selection takes place in the first or the second step of each evolutionary event. For the local and global death-birth processes ($\dB$, $\Db$) the fixation probability is lower than on a complete graph, as shown by the green and blue lines in Fig.~\ref{fig:FixProbS_R}. Conversely, both \Bd ~and \bD ~show a higher fixation probability than on complete graphs (red and purple lines).

In the limit of fast flows, however, the outcome of evolution is mostly determined by whether selection is global or local, and not by the order of the reproduction and removal events (birth-death vs. death-birth). Specifically, when selection acts locally the fixation probability of the mutant is lower than on a complete graph (purple and blue lines). In contrast, when selection is global the fixation probability is the same as on a complete graph (red and green lines).

\medskip
These observations indicate unique responses of the fixation probability to the flow speed for the different processes. For the \Db ~process (continuous green line in Fig.~\ref{fig:FixProbS_R}) the mutant's probability of success increases with the speed of the flow.
For the \Bd ~process (continuous red line), the fixation probability decreases with increased flow speed, but is always greater than or equal to the one on a complete graph, $\phi\geq\phi_{\text CG}$. In contrast, the fixation probability for a \dB ~process (dashed blue line) is always smaller than $\phi_{\text CG}$. Finally, for the \bD ~process (dashed purple line) the fixation probability is higher than on a complete graph when the flow is very slow, but decreases at higher flow speeds and eventually becomes lower than on the complete graph. The \bD ~process is the only case in which we observe a transition from amplification to suppression of selection (relative to the complete graph) as the flow speed is increased.

\medskip
In order to gain some insight into these observations, we first describe the dynamics in the limit of fast flows, summarising the results of ref.~\cite{Herrerias2018}. Then we discuss the no-flow limit, and subsequently the transition between the two extremes, at intermediate flow speeds.

\subsubsection*{Fast-flow limit: evolution of well-stirred populations}
When the flow is sufficiently fast the probability that any two particles are neighbours at the time of an evolutionary event is the same, irrespective of whether they were neighbours at the previous event or not\cite{Herrerias2018}. In global processes selection takes place when the first individual is chosen, i.e., competition acts amongst the whole population. Then, a second individual is picked at random from the neighbours of this first individual. Since any individual is equally likely to be neighbours of the individual selected in the initial step, the second, random pick, is equivalent to a random pick from the entire population. Therefore, in the limit of fast flows the fixation probability of global processes coincides with the one on complete graphs, as observed in Fig.~\ref{fig:FixProbS_R}.

For local processes, on the other hand, in each evolutionary event the first individual is chosen at random from the entire population, irrespective of fitness. Competition then takes place between the neighbours of this individual. Although all members of the population are equally likely to be part of this neighbourhood, at any one time the group of neighbours is a random subset of the population. This subset may not reflect the composition of the population as a whole, which can be shown to lead to suppression of selection\cite{Herrerias2018}. We briefly illustrate this for the case of a very small interaction radius; the majority of individuals then have at most one neighbour at any given time. Since this neighbour is the only contestant in local selection, fitness is irrelevant. Therefore, as the interaction radius becomes small the fixation probability of the mutant approaches the limit of neutral selection. When the interaction radius is large, however, it is more likely that the group of neighbours is large as well, and that population-wide frequencies are accurately represented. Therefore, the suppression effect relative to the complete graph is reduced. If the interaction range is so large that all individuals are connected with all other individuals at all times, a complete interaction graph is recovered.

Analytical results can be obtained for all four processes in the limit of very fast flows\cite{Herrerias2018}. Predictions from this theoretical approach are shown as filled circles on the right edge of Fig.~\ref{fig:FixProbS_R}.

\subsubsection*{No-flow limit: evolution on static heterogeneous graphs}
On the left-hand side of Fig.~\ref{fig:FixProbS_R} the flow is so slow that the evolutionary dynamics effectively take place on fixed graphs. Evolutionary processes on static graphs have been widely discussed in the literature (see e.g. \cite{Nowak2010, Perc2013, Shakarian2012} and references therein). The focus is often on characterizing specific graphs or graph structures, which either amplify or suppress selection\cite{Hindersin2016,Giakkoupis2016,Adlam2015}. Notably, the authors of ref.\cite{Hindersin2015} report that most undirected graphs amplify selection for birth-death processes, but suppress selection for death-birth processes. However, these findings are only given for relatively small networks, and only for processes in which selection acts in the reproduction step (\Bd ~and \dB).  

\begin{figure}[b]
	\center
	\includegraphics[width=0.45\textwidth]{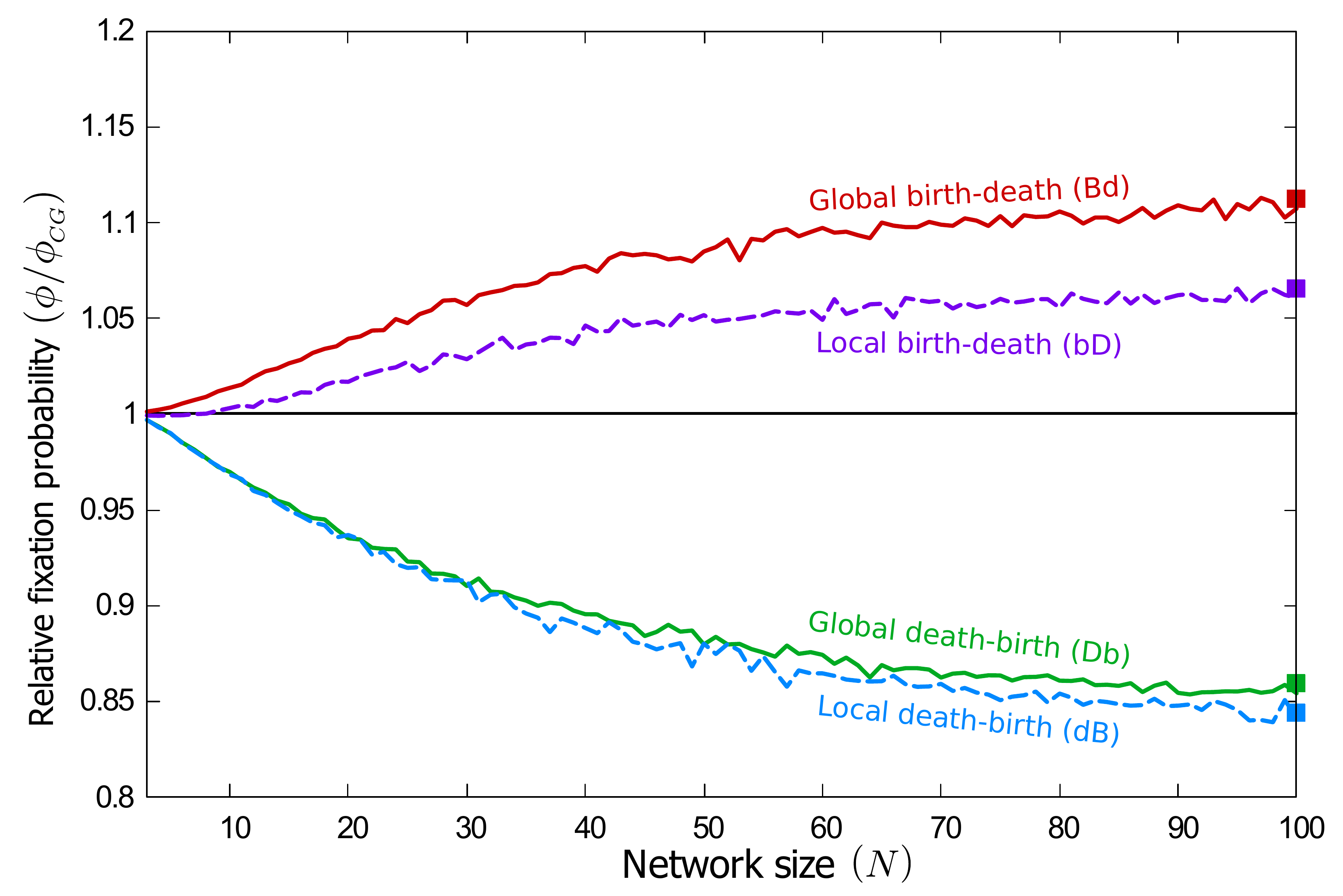}
	\caption{\textbf{Fixed heterogeneous graphs amplify selection for birth-death processes and suppress it for death-birth processes.} 
	The figure shows the fixation probability of an invading mutant ($\phi$), averaged over static CRGGs. Data is shown relative to the corresponding fixation probability on a complete graph ($\phi_{CG}$).
	Regardless of the population size, selection is amplified for \Bd ~and \bD ~processes, and suppressed for \Db ~and \dB ~processes.
				}
	\label{fig:RandDomFix}
\end{figure}

In order to obtain a more complete picture, we measured in simulations the fixation probability of a single mutant on networks of different sizes, averaged over different static heterogeneous graphs. Each graph is generated by placing individuals at random in the spatial domain (see Sec.~\ref{sec:Methods}), resulting in a random geometric interaction graph. It is possible that a graph generated in this way consists of several disconnected components. In the absence of flow, the mutant then cannot reach fixation. We therefore restrict simulations to graphs with a single connected component and henceforth use the term \textit{connected} random geometric graphs (CRGGs). We present results for the different evolutionary processes as a function of the size of the graph in Fig.~\ref{fig:RandDomFix}.

The data shows that the average fixation probability of a single mutant on CRGGs is higher than on the complete graph for birth-death processes, $\phi\geq\phi_{\text CG}$. For death-birth processes, on the other hand, $\phi\leq\phi_{\text CG}$. This is in line with the results reported in ref.\cite{Hindersin2015} for small graphs. The data in Fig.~\ref{fig:RandDomFix} confirms that the amplification of selection (for birth-death processes) or suppression (for death-birth processes) is present regardless of the size of the network, if an average over many graphs is taken. The rightmost data points in Fig.~\ref{fig:RandDomFix} correspond to a population of the same size as the one in Fig.~\ref{fig:FixProbS_R}.

Intuition regarding the amplification or suppression of selection on static networks can be gained by studying the connectivity of the initial mutant (see refs.\cite{Antal2006,Sood2008,Broom2011,Maciejewski2014,Tan2014}). For death-birth processes, these studies find that the success of an advantageous mutant increases with its degree; for birth-death processes, its success decreases with connectivity. This can be understood in the following way: In each evolutionary event two individuals are chosen, the first from the entire population, and the second as a neighbour of the first. The degree of an individual does not affect its chances of being chosen in the first step, irrespective of whether selection acts in this step or not. However, the probability of being neighbours with the initial individual is higher for well connected individuals than for individuals with a low degree. Under birth-death processes, higher connectivity of the mutant therefore results in a higher chance of being replaced. For death-birth processes it results in a higher chance of reproduction.

\begin{figure}[t]
	\center
	\includegraphics[width=0.45\textwidth]{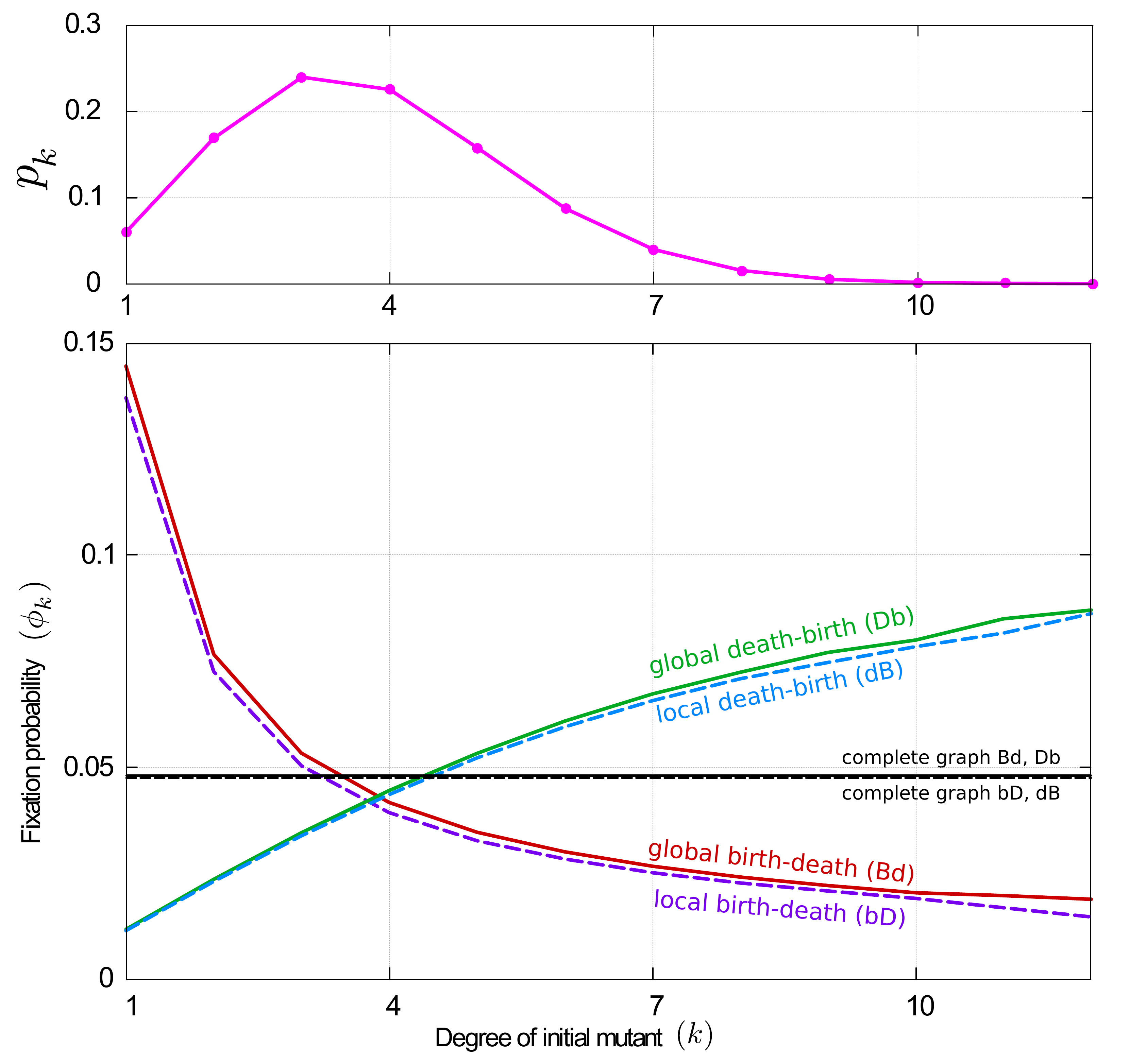}
	\caption{\textbf{Significance of the degree of the initial mutant.} The upper panel shows the degree distribution, $p_k$, of the ensemble of connected random geometric graphs (CRGGs), obtained by placing $N=100$ individuals into the spatial domain $0\leq x,y\leq 1$ with uniform distribution, and using an interaction radius $R=0.11$ and periodic boundary conditions.
	The lower panel shows the fixation probability obtained from simulating the evolutionary process on these graphs, as a function of the degree of the initial mutant. For the two death-birth processes the mutant's success is below the one on a complete graph if its degree is low, and above $\phi_{CG}$ at high connectivity. The reverse is found for the two birth-death processes. Data points have been connected as a visual guide.}
	\label{fig:MutPos0}
\end{figure}

In the literature these predictions have been tested for \Bd ~and \dB ~processes \cite{Antal2006,Sood2008,Broom2011,Maciejewski2014,Tan2014}. In the lower panel of Fig.~\ref{fig:MutPos0} we verify that the argument extends to all four evolutionary update rules defined above. We show, for CRGGs, the fixation probability of a mutant, $\phi_k$, as a function of its degree, $k$. For the global and local death-birth processes (\Db, \dB) the mutant's success is lower than on a complete graph when the mutant is sparsely connected, but larger if it is highly connected; the reverse is found for global and local birth-death processes (\Bd, \bD). These observations are consistent with the above reasoning.

\medskip
In our model, the initial mutant is chosen uniformly at random from the members of the population. The probability that it has degree $k$ is thus determined by the degree distribution of CRGGs. We write $p_k$ for the probability of a random node to have degree $k$ in such a graph, and show the degree distribution for networks of size $N=100$ in the upper panel of Fig.~\ref{fig:MutPos0} for illustration. The overall probability of fixation of a single mutant is then $\phi=\sum_k p_k\phi_k$. Fixation probabilities obtained in this way are shown as square markers in Fig.~\ref{fig:FixProbS_R}~and~\ref{fig:RandDomFix}.
The results  reproduce the amplification and suppression of selection (for birth-death or death-birth processes, respectively) in the limit of slow flows. We attribute quantitative differences between the markers and lines in Fig. \ref{fig:FixProbS_R} to effects of the non-zero flow and to the difference in initial conditions; the data shown as lines is obtained from simulations of slowly flowing populations in which the initial graph may consist of more than one component. In  Sec.~\ref{sec:InitialPositions} we will discuss the difference due to initial positions in more detail.

\subsubsection*{Transition between fast-flow and no-flow limits}
As seen above, the outcome of evolution in rapidly stirred populations is very different to that on static interaction graphs. With fast flows, local competition leads to suppression of selection; on the other hand, the success of a mutant is the same as on a complete graph if selection is global. When there is no flow, the order of the birth and death events in the evolutionary process is crucial. In this case, selection is amplified for birth-death processes and suppressed for death-birth processes. At intermediate flow speeds, a crossover between these two regimes is seen. We will now discuss this transition in more detail.

On fixed heterogeneous graphs, the degree of the initial mutant determines whether its chances of success are greater or smaller than on a complete graph. In the presence of flow the interaction network constantly changes, and the number of neighbours of any one individual thus varies over time. Classifying a member of the population as highly or poorly connected is then at best possible over limited time windows. If the flow is slow relative to evolution, many evolutionary events occur in such a time window, and the evolutionary dynamics can conclude before the degrees of nodes undergo significant changes. Therefore the amplification or suppression effect due to the degree of the mutant can still be observed. For faster flows, however, the interaction network changes so quickly that there is no clearly defined notion of a degree of an individual on the time scale of evolution. The amplification or suppression effect set by the initial heterogeneous network is then washed out.

At very fast flow speeds, the set of neighbours of the individual chosen in the first step of an evolutionary update effectively becomes a group sampled at random from the entire population. Therefore, suppression of selection sets in for local processes. The fixation probability of global processes, on the other hand, approaches the one on a complete graph, as described previously.

The main effects leading to the transition between the no-flow and the fast-flow limits are thus the increasing variability (over time) of the degree of individuals, and the random sampling of the group of individuals taking part in evolutionary events. As a result of these two mechanisms, in Fig.~\ref{fig:FixProbS_R} we see a smooth transition between the two limits. The different responses of the fixation probability to the speed are a consequence of the limiting behaviours for very slow and very fast flows.

Although it is not immediately transparent from the results in Fig.~\ref{fig:FixProbS_R}, the flow has further effects on the evolutionary process. For example, it removes the influence of the initial positions of the individuals in space. Another important feature, particularly at intermediate flow speeds, is that the evolutionary process takes place on slowly changing heterogeneous graphs. The dynamic network constantly splits into disconnected components, which later merge and form new components. This fragmentation promotes the formation of `clusters' --- groups of nodes which are of the same species. This gives rise to further amplification or suppression effects, depending on the details of the evolutionary mechanics. In the following section, we explore these effects further.

\subsection{Effects of the initial positions of individuals}\label{sec:InitialPositions}
Our model describes a population in constant motion. It is then natural to assume that the positions of the individuals at the time the initial mutation occurs is drawn from the stationary distribution of the flow. For the periodic parallel shear flow this is the uniform distribution, used as an initial condition in the previous section. However, exploring different starting positions allows us to gain further insight into the effect of the flow on fixation probabilities.  

\subsection*{Connected random geometric graphs (CRGGs)}
The data shown as lines in Fig.~\ref{fig:FixProbS_R} was obtained from simulations with random initial positions (RGGs) and non-vanishing flows. For this setup the interaction graph may not be connected, but fixation or extinction will still occur, provided there is non-zero flow. In order to explore the no-flow limit, in Figs.~\ref{fig:RandDomFix}~and~\ref{fig:MutPos0} we focused on static heterogeneous graphs instead; studying fixation in the strict absence of flow only makes sense when the interaction graph consists of one single connected component, and so we restricted the discussion to connected random geometric graphs (CRGGs).
As a result, comparison with the data in Fig.~\ref{fig:FixProbS_R} is difficult; in particular, we note the quantitative differences between the square markers, obtained from static connected graphs, and the limiting values of the data shown as lines in Fig.~\ref{fig:FixProbS_R}, obtained from slowly moving populations started from RGGs.

\begin{figure}[t]
	\center
	\includegraphics[width=0.65\textwidth]{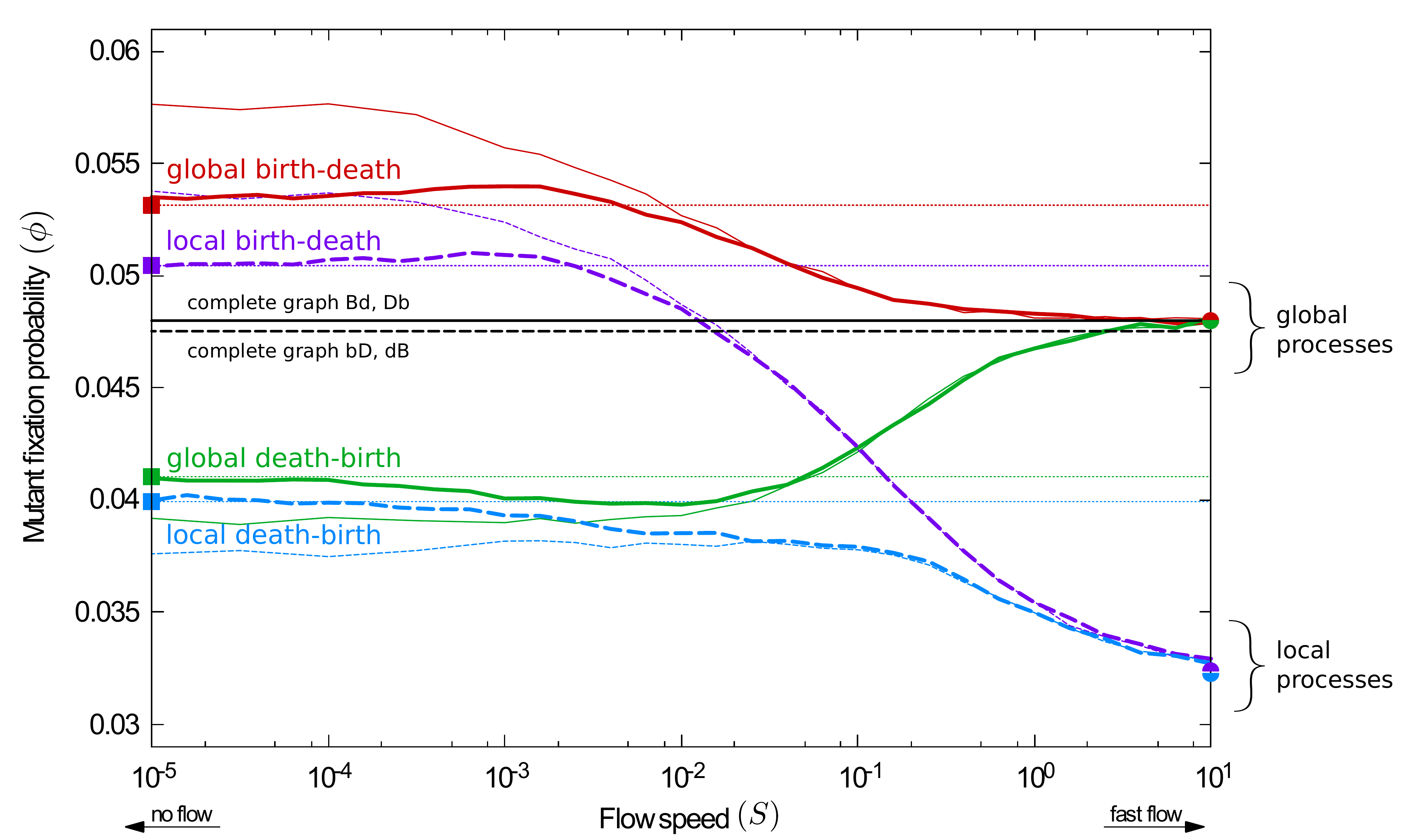}
	\caption{\textbf{Comparison of fixation probability for simulations started from unrestricted and connected random geometric graphs (RGGs and CRGGs, respectively).} 
The fixation probability as a function of the flow speed is shown as thick lines for simulations started on connected graphs; thin lines are for unrestricted initial positions (some of this data is also shown in Fig.~\ref{fig:FixProbS_R}). Square markers indicate the fixation probabilities on \textit{static} CRGGs; see text for details. The fixation probability on complete graphs is shown for reference. A minimum of $\phi$ is found for the \Db ~process; maxima are discernible for \Bd ~and \bD ~when the dynamics are started from connected graphs. The effect of amplification/suppression of selection at slow flow speeds is more pronounced for simulations initialized from RGGs than from CRGGs.
	}
	\label{fig:FixProbS_C}
\end{figure}

For comparison, we show data obtained from mobile populations, but started on CRGGs, in Fig.~\ref{fig:FixProbS_C}. The limiting values of the fixation probabilities for very slow flows (end of the tick lines on the left-hand side of the figure) now agree quantitatively with those obtained from static CRGGs (square markers).

The simulation data from Fig.~\ref{fig:FixProbS_R}, from simulations with unrestricted random initial positions, is also shown in Fig.~\ref{fig:FixProbS_C} (thin lines). If the flow is sufficiently fast, initial conditions are immaterial. On the contrary, for slow flows the fixation probability, $\phi$, for simulations started from unrestricted random graphs is different from that for connected initial conditions. For birth-death processes, $\phi$ is greater for the unrestricted case than for the connected one. The opposite is observed for death-birth processes. This indicates that the initial condition can have a significant effect on the outcome when then flow is slow.

\begin{figure}[t]
	\center
	\includegraphics[width=0.65\textwidth]{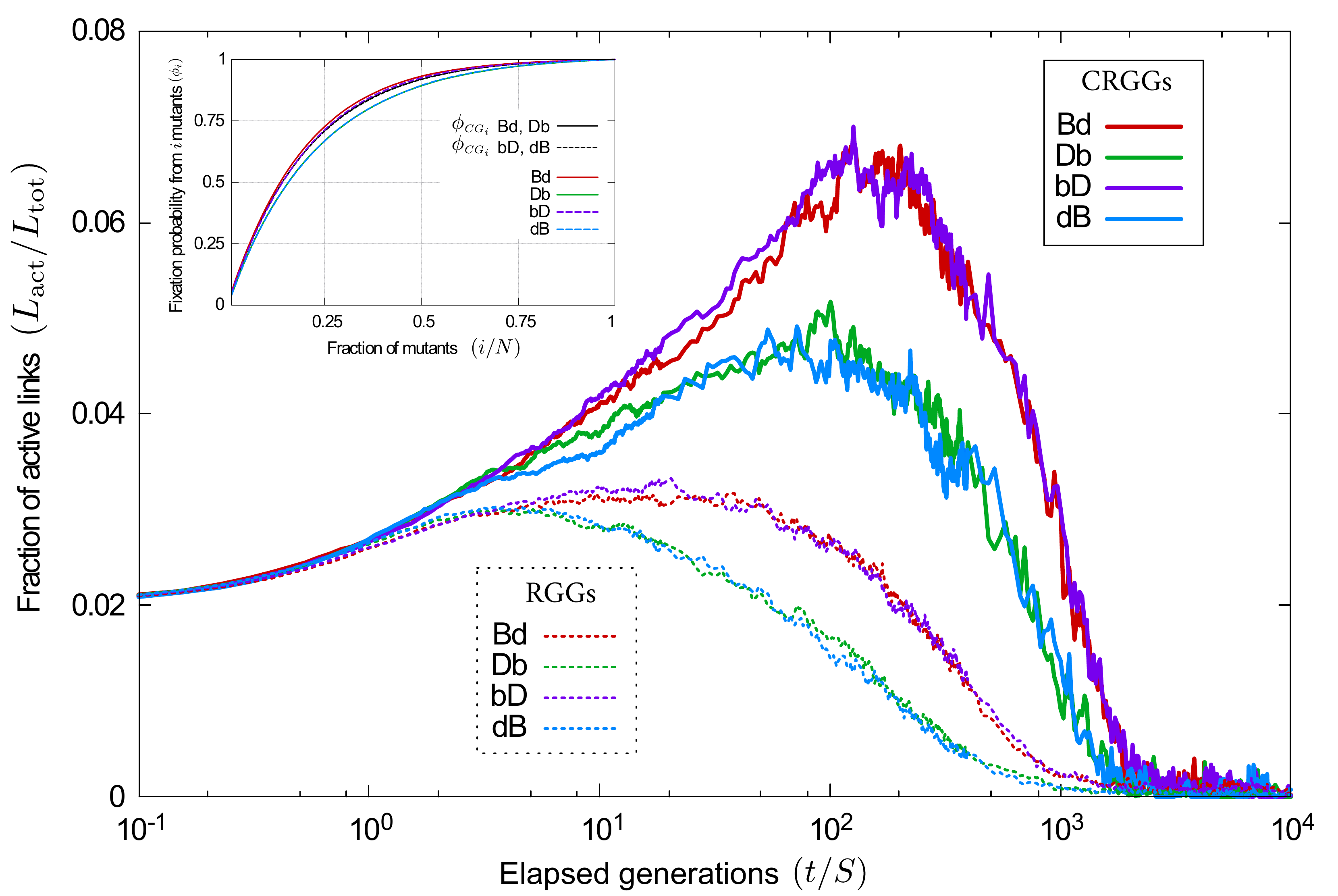}
	\caption{\textbf{Fragmented initialization promotes the formation of clusters.} 
The main panel shows the average proportion of active links as the evolutionary dynamics proceed. Thick lines correspond to simulations started from connected graphs (CRGGs); thin dotted lines to simulations initialized from unrestricted random positions (RGGs). The fraction of active links is lower for RGGs, regardless of the evolutionary process. \textbf{Inset:} Fixation probability of the mutant species, once there are $i$ mutants in the population. When mutants are a minority, a small increase in their frequency greatly increases their fixation probability. Conversely, reducing their numbers when they are a majority has only minor effects on their chances of success. Simulations in the inset are initialized from CRGGs.
				}
	\label{fig:ActiveLinks}
\end{figure}

\medskip
As briefly mentioned before, the fragmented nature of the unrestricted setup can isolate groups of nodes from the rest of the population. As the evolutionary dynamics proceed, this promotes the formation of clusters, i.e. parts of the graph in which all individuals are of the same species.
The degree of clustering can be quantified through the fraction of active links in the network, that is, the proportion of links between mutants and wildtypes among all links in the graph, $L_{\rm act}/L_{\rm tot}$. A small fraction of active links is an indicator of clustering. We show measurements of the fraction of active links in Fig.~\ref{fig:ActiveLinks} for both unrestricted and restricted random initial conditions (thin dotted lines and thick continuous lines, respectively). The data indicates that the fraction of active links is significantly larger when simulations are initialised on CRGGs than when started on RGGs.

The amplification or suppression of selection (for birth-death and death-birth processes, respectively) can then be supported by a similar argument to the one presented for the degree of the initial mutant. A smaller number of active links has the same effect as poor connectivity of the initial mutant; it does not affect the probability that the individual chosen in the first step of the evolutionary process is a mutant or a wildtype, but it reduces the probability that the individual chosen in the second step is of the opposite species (see also Sec.~\ref{app:WheelGraphs} of the Supporting Information).

In the early stages of the evolutionary process mutants are a minority, and are therefore less likely to be chosen in the initial step. A large number of active links then increases the chances that the neighbour of the initial individual is a mutant. Under birth-death processes this means that mutants are more likely to die; for death-birth processes they have more opportunities to reproduce. Therefore, a connected initial configuration (CRGGs), leading to a larger fraction of active links than arbitrary RGGs, reduces the fixation probability of a mutant under birth-death processes, and increases it for death-birth processes. This is in line with the results on the left-hand side of Fig.~\ref{fig:FixProbS_C}; the fixation probabilities for RGGs (dotted lines) are higher than their counterparts on CRGGs for birth-death processes (red and purple lines), but lower for death-birth processes (green and blue lines).

This argument is only valid when mutants are less abundant than wildtypes. The effect is reversed at later stages of the evolutionary process (if mutants become a majority). However, the results presented in Fig.~\ref{fig:FixProbS_C} suggest that there is a net advantage for the mutant in having fewer active links, for birth-death processes, or in having increased inter-species connectivity, for death-birth mechanics.
The inset in Fig.~\ref{fig:ActiveLinks} helps to understand this further. It shows the conditional fixation probability of the mutant species, given that a state with $i$ mutants has been reached. The shape of the curves indicates that increasing the number of mutants in the population has stronger repercussions on the fixation probability when mutants are a minority ($i/N\leq0.5$) than when they are the majority ($i/N\geq0.5$).
For death-birth processes, the selective effect due to increased active links drives the population composition to states with approximately equal frequencies of the two species. However, the mutants have more to gain (in terms of fixation probability) when their numbers are small than what they may lose when they are abundant. For birth-death processes, on the other hand, a large number active links acts in the opposite way; it hinders the spread of the mutant species when they are a minority and encourages it once they are abundant. Since more is lost in the early invasion than what can be gained at later stages, the overall fixation probability is lower than when there are fewer active links. The net effect of fragmentation (i.e., a reduced number of active links) is therefore amplification of selection for birth-death processes, and suppression for death-birth update rules.

\medskip
The amplification/suppression effect caused by the fragmented nature of the network can also be noticed at intermediate flow speeds. In this regime, the flow is sufficiently fast to disrupt the initial network structure before the evolutionary process reaches its conclusion (fixation or extinction of the mutant); disconnected components then develop. At the same time the flow is also slow enough to allow the formation of organised clusters of mutants and wildtypes through the evolutionary dynamics. Indeed, for simulations started on connected graphs a minimum in the fixation probability as a function of the flow speed is discernible for the \Db ~process (thick green line in Fig.~\ref{fig:FixProbS_C}), and we also notice a shallow maximum for the \Bd ~and \bD ~processes (thick red and purple lines, respectively). The fragmentation from an initially connected network increases the fixation probability for birth-death processes and decreases it for death-birth processes. Movement of the population, and the resulting mixing between evolutionary events counteracts this amplification or suppression, driving fixation probabilities to their fast-flow limits. The balance of these two effects leads to the extrema in Fig.~\ref{fig:FixProbS_C}.

\subsection*{Square lattice}
Regular lattices are particularly convenient for the study of fixation probabilities. The nodes are distributed equidistantly in space, and they all have the same number of neighbours. This means that analytical results can be obtained in the absence of flows. For example, the isothermal theorem\cite{Lieberman2005} applies; the fixation probabilities of the global birth-death and death-processes are the same as those for complete graphs; only small deviations from $\phi_{CG}$ are expected for local-selection processes\cite{Kaveh2015}.

In order to relate the success of mutants in populations advected by flows to these benchmark results, we show the outcome of simulations in which individuals are initially placed on the nodes of a regular lattice in Fig.~\ref{fig:FixProbS_L}. Broadly, three different regimes can be distinguished:

\begin{figure}[t]
	\center
	\includegraphics[width=0.63\textwidth]{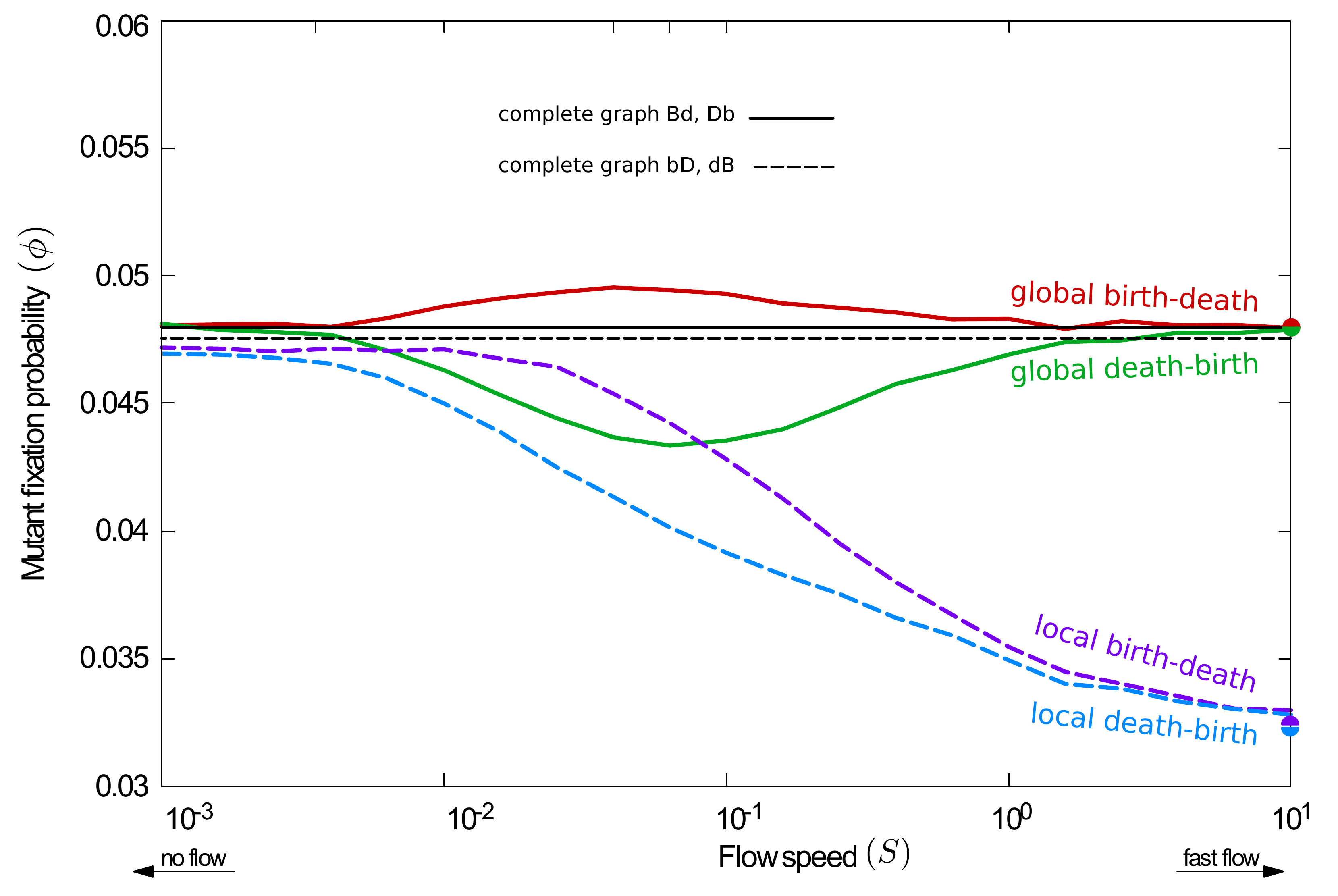}
	\caption{\textbf{Fixation probability at different flow speeds for simulations started from a square lattice.} 
For the global death-birth process a minimum of fixation probability is found at intermediate flow speeds; conversely, the global birth-death process shows a maximum. For the local processes no extrema are found; instead varying the flow speed interpolates monotonously between the behaviour on fixed lattices and the limit of fast flows.
	}
	\label{fig:FixProbS_L}
\end{figure}

\medskip
{\em Quasi-isothermal regime.} On the left-hand side of Fig.~\ref{fig:FixProbS_L} (slow flows) fixation probabilities for all processes are approximately the same as on complete graphs. This is to be expected; in the limit of slow flows the evolutionary process concludes before the lattice structure is modified. The interaction network remains regular and, in line with the isothermal theorem, the fixation probability for global processes (continuous lines) is the one known from complete graphs; results for local processes (dashed lines) only differ slightly from $\phi_{CG}$.

With the periodic parallel shear flow, this agreement extends to slow, but non-vanishing flows. As described in Sec.~\ref{sec:Methods}, during the first half of each period the flow moves the particles only vertically, with velocities dependent on their horizontal position. This means that some elements of the initial lattice remain intact; for example initial `columns' of individuals (those with the same horizontal coordinate) move jointly. There is then only limited variation in the degree of the nodes in the network, and the interaction graph remains nearly regular. If fixation or extinction occurs before the flow disrupts this {quasi-isothermal} structure, the predictions of the isothermal theorem remain a good approximation. The flow speed above which this is no longer the case can be estimated from a comparison of the the time until the lattice structure is disrupted and the time-to-fixation; see Sec.~\ref{app:Timescales} of the Supporting Information for further details. 
\medskip

{\em Intermediate regime.} At intermediate flow speeds the fixation probability for the \Bd ~process exhibits a maximum; a minimum is found for the \Db ~process. These features can be related to the amplification or suppression effects on heterogeneous graphs, discussed in the previous sections. For intermediate flow speeds, the individuals' motion is fast enough to distort the initial lattice structure before the evolutionary process concludes. On the other hand the flow is also sufficiently slow so that evolution has time to organise in clusters on the heterogeneous interaction network. Effectively evolution takes place on a slowly moving heterogeneous graph. This heterogeneity, in conjunction with the clustering of species, leads to amplification of selection for birth-death processes and suppression for death-birth processes. When selection is local this merely accelerates or delays the approach to the behaviour on complete graphs.  When selection is global, however, the minimum (for \Db) and maximum (for \Bd) are generated. A rough estimate for the flow speed at which the extrema are seen can be obtained by comparing the time-to-fixation of the mutant species with the network renewal time; details can be found in Sec.~\ref{app:Timescales} of the Supporting Information.

\medskip
{\em Fast flow.} In this regime the positions of individuals in space at each evolutionary event are essentially random, and the set of neighbours of any one particle is uncorrelated from an evolutionary event to the next one. The population is then `well-stirred', and the analytical predictions from Ref.~\cite{Herrerias2018} apply.

\section{Discussion}\label{sec:Discussion}
We studied evolutionary dynamics in populations immersed in flows. In computer simulations, we measured the effect that the speed of the motion has on the success of an invading mutant, and found that different evolutionary processes show distinct responses of the fixation probability to the flow speed. 
Our results highlight the importance of including motion in the modelling of evolutionary dynamics. Just as population structure can generate amplification or suppression of selection, we find that the flow can act against or in favour of mutant invasion. While the models we study are stylised, we can identify general emerging principles. For instance, for the majority of evolutionary processes we observe a decrease in fixation probability when populations are in motion. This observation could be useful, for example, in industries where mutations are detrimental for the desired product but beneficial to the mutant, such as in microalgae, bacteria, fungi and yeast, relevant for the production of biodiesel\cite{Ma1999,Chisti2007a,Jeon2010a,Meng2009a}. Another example are the features we found to dominate fixation probability in the limits of very slow or very fast flows. If populations are mostly static in an experiment, our results indicate that whether selection acts locally or globally is a more important factor than the order of birth and death events. If an experiment involves populations in motion, on the other hand, careful consideration has to be given whether to use a birth-death or a death-birth process as a model, and it is less important in what step of evolutionary events competition takes place.

We hypothesise that the characteristic responses to flow may be used as an aid for choosing the most adequate update mechanism to model a given biological system. Despite the fact that direct measurements of the success of a specific mutation are not necessarily easy to perform, there is experimental evidence of differences in fixation probabilities in static and in stirred populations\cite{Kerr2002,Habets2006,Perfeito2008a}. In these studies, cultures of \textit{E. coli} were grown in a continuously stirred liquid medium, on Petri dishes, mixed every 24 hours, and on static Petri dishes. The structure and cluster formation of the cultures were found to have different dynamics under the different mixing conditions. The authors of ref.~\cite{Perfeito2008a}, for example, find that the ability to adapt, as measured by reproduction rates, is greater in the continuously-stirred case than in the case of only occasional mixing. This suggests a lower fixation probability in the slowly moving medium. Comparing this with our results, we speculate that a \Db~process might best describe this biological system.

It is appropriate to briefly comment on the limitations of our study. For example, we focused on the periodic parallel shear flow in our simulations. However, we note that most features of the amplification or suppression of selection are not due to particulars of the flow field. Instead they arise from the mixing of the population and the heterogeneity of the interaction network. Both of these features can be expected in most real flows, and we believe that the essence of our findings is relevant beyond the exemplar of the shear flow. This is supported by observations in our earlier work\cite{Herrerias2018}, in which we obtained analytic results for limit of fast flows and demonstrated that these predictions are independent of many details of the flow field. Our study is also limited to frequency-independent selection; natural extensions would include more complex fitness functions to better model the experimental situation in ref.\cite{Habets2006}, where frequency-dependent fitness was identified for completely static conditions. We are aware that our simulations are for relatively small populations; this is due to computational costs associated with numerical experiments on a larger scale. Further work may also be necessary to relax assumption of a fixed population size. This may be useful to explore the effects of demographic stochasticity. On the other hand, dilution techniques or resource-limited environments can be used in experiments to keep the population approximately constant without significantly modifying the mutants' chances of success\cite{Wahl2002}. 

Recent advances in technology make direct measurements of the fixation probability of a specific mutation feasible\cite{Patwa2008}. We believe that this, together with computational studies of different evolutionary models in varying conditions, can open up promising routes to more informed choices of evolutionary mechanics for systems in evolutionary biology.


\section*{Acknowledgements}\vspace{-.2cm}
FHA thanks Consejo Nacional de Ciencia y Tecnolog\'ia (CONACyT, Mexico) for support. VPM acknowledges  financial support by CRETUS strategic partnership (ED431E2018/01), co-funded by the ERDF (EU).
We would also like to thank Joseph W. Baron for his helpful comments.

\vspace{-.2cm}
\section*{Author contributions statement}\vspace{-.3cm}
FHA carried out the simulations and analytical calculations. All authors contributed to designing the research, to analysing the data and to writing the paper.

\vspace{-.3cm}
\section*{Additional information}\vspace{-.3cm}
We declare no competing interests.


%


\label{LastPageDoc}		

\clearpage
\clearpage
\setcounter{section}{0}		
\setcounter{page}{1}		
\setcounter{equation}{0}	
\setcounter{figure}{0}		
\renewcommand{\thesection}{S\arabic{section}} 		
\renewcommand{\thepage}{S\arabic{page}} 			
\renewcommand{\theequation}{S\arabic{equation}}  	
\renewcommand{\thefigure}{S\arabic{figure}}  		

\begin{center}
\vspace*{3cm}
{\LARGE \bf Supporting Information}\\ \vspace{.2cm}
{\large \bf \TITLE}\\
\vspace{0.6cm}
{\large Francisco Herrer\'{i}as-Azcu\'{e}${}^{\rm 1}$, Vicente P\'{e}rez-Mu\~{n}uzuri${}^{\rm 2}$  and Tobias Galla${}^{\rm 1}$}\\
{\tt francisco.herreriasazcue@manchester.ac.uk, vicente.perez@cesga.es, tobias.galla@manchester.ac.uk}\\
\vspace{1em}
${}^{\rm 1}$Theoretical Physics, School of Physics and Astronomy, \\ The University of Manchester, Manchester M13 9PL, United Kingdom\\
\vspace{1em}
${}^{\rm 2}$Group of Nonlinear Physics, Faculty of Physics, \\ University of Santiago de Compostela E-15782 Santiago de Compostela, Spain

\end{center}

\section{Identification of relevant timescales for simulations started from regular lattices }\label{app:Timescales}
In Sec.~\ref{sec:InitialPositions} of the main text we discuss the behaviour of the population when the initial condition is a regular lattice. We identify  three different regimes: fast flows, intermediate flow speeds, and the quasi-isothermal regime. Here we discuss how these regimes can be identified from the simulations, and show how the time scales for the network renewal and quasi-isothermal regimes can be obtained. Simulations in this section were all initialized from a square lattice. The flow is the periodic parallel shear flow, with parameters as given in the main text.

\subsection{End of quasi-isothermal regime}\label{sec:eqir}
\begin{figure}[htb!!]
	\center
	\includegraphics[width=0.6\textwidth]{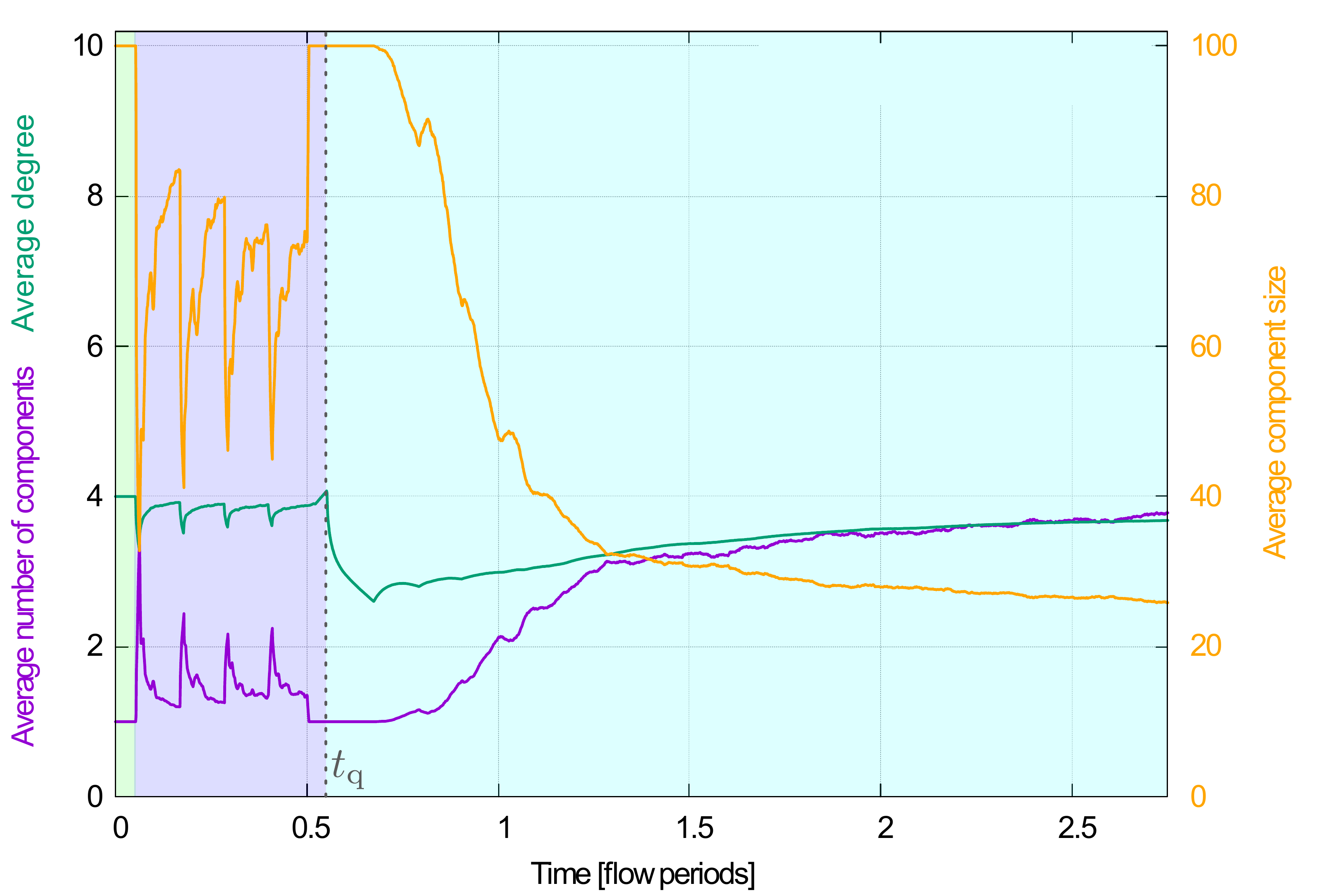}
	\caption{\textbf{Component formation as flow destroys initial lattice configuration.} 
The average number of components (purple) and the average degree (green) are plotted on the left axis; the average component size (orange) is plotted on the right axis. The three phases of the motion described in Sec. \ref{sec:eqir} of the Supplementary Information are shaded in different colours. The grey dotted line at $t_{\rm q}$ marks the end of the phase in which the graph is quasi-isothermal.}
	\label{fig:FlowDom}
\end{figure}

Fig.~\ref{fig:FlowDom} shows the average number of components in the interaction graph, the average size of each component, and the average degree in the network as a function of time.

We observe three different phases:
\begin{enumerate}[label=(\roman*),nolistsep]
	\item At first, there is a very short interval ($t\lesssim 0.05$), in which there is only one component of size $N$. The lattice interaction graph is still intact.
	\item Next, for $0.05\lesssim t\lesssim 0.55$ we see an oscillating number of components; this corresponds to a period in which movement is only vertical (for the most part). Links between individuals with the same horizontal coordinate, i.e., within a `column' of the original lattice, are not modified. The interaction graph is nearly regular.
	\item Following this, we see a sharp decline of the average degree. The network cannot be considered isothermal any more, as fragmentation into heterogeneous separate components has begun. All measured quantities approach their stationary asymptotic values (the figure shows an average over multiple runs). Initial positions are washed out and the individuals take random positions in space.
\end{enumerate}

\medskip
We refer to the two initial stages (i) and (ii) as the `\textit{quasi-isothermal}' period. If fixation (or extinction) is reached in most runs during this initial period, we expect fixation probabilities close to the one on the complete graph. For the model parameters in our simulations, phase (ii) ends at $t_{\rm q}\approx 0.55$, shortly after the first half period of the flow ($T/2=0.5$). This is marked with a dotted grey line in Fig.~\ref{fig:FlowDom}. 
We note that the notion of the quasi-isothermal regime relies on the conservation of elements of the regular lattice; this may not be the case in other flows, for example if the motion of particles is not strictly vertical or horizontal. However, regardless of the type of flow, a critical flow speed can be found, below which the evolutionary process concludes before the lattice structure is significantly modified. Therefore, distinction of the regimes remains broadly valid.

\subsection{Network renewal time}
\begin{figure}[htb!!]
	\center
	\includegraphics[width=0.5\textwidth]{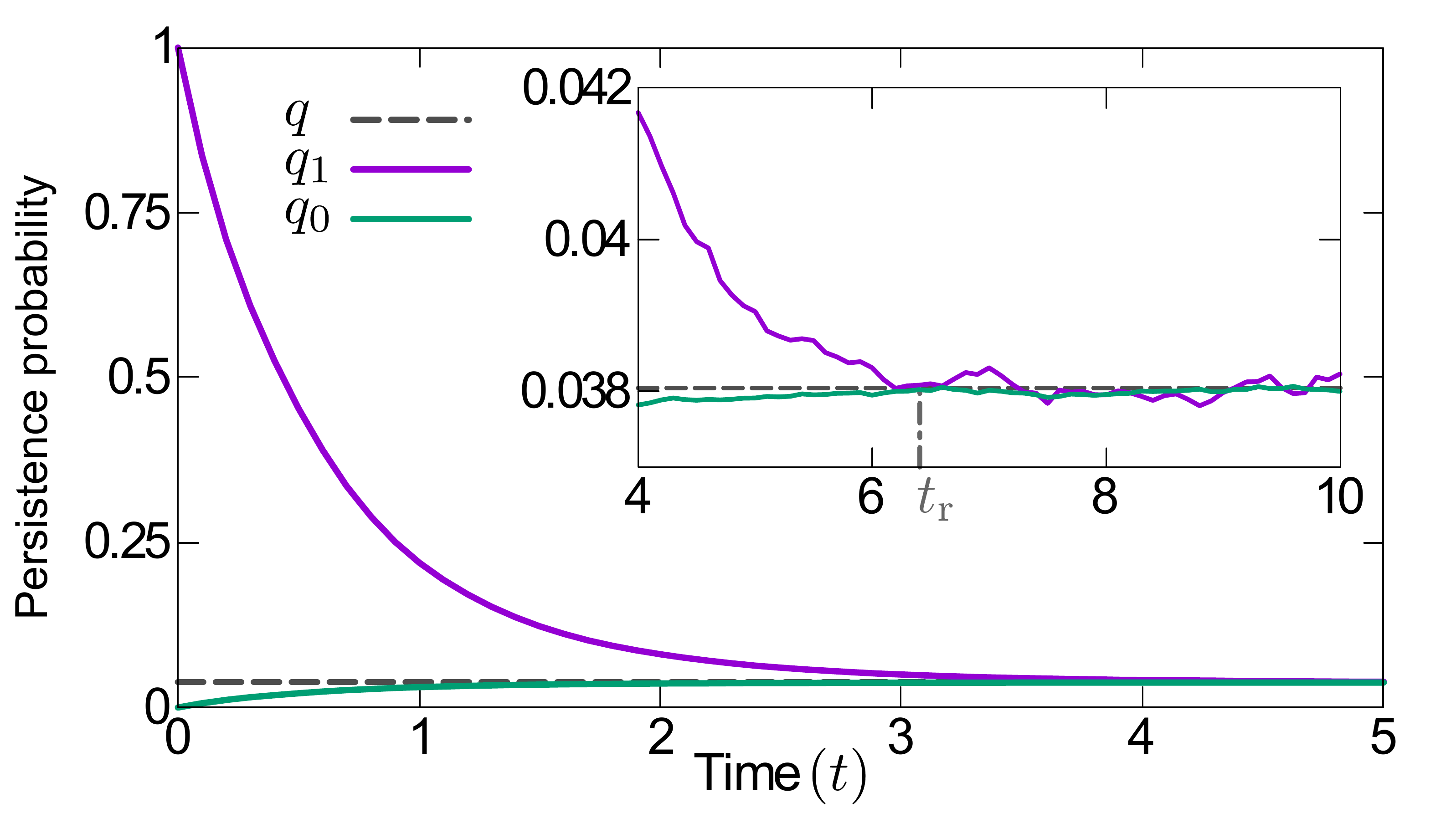}
	\caption{\textbf{Network-renewal time measured from the persistence of links.}  Continuous lines show the probability that two nodes, connected (purple) or disconnected (green) at $t_0$ are still connected/disconnected at time $t_0+t$. The dashed grey line shows the asymptotic value; the time needed for both probabilities ($q_1$ and $q_0$) to reach this value is the time it takes to renew the network. Both quantities are within $0.1\%$ of their asymptotic value for the first time at $t_r\approx6.4$, marked by a vertical dash-dotted grey line in the inset.
	}
	\label{fig:LinkProb}
\end{figure}

The time needed for neighbourhoods of individuals to lose correlation defines the \textit{renewal} time, $t_{\rm r}$. To measure this we have looked at the persistence of links in the network.  As in ref.\cite{Herrerias2018}, we consider the probability that two nodes, connected at time $t_0$, are still connected at time $t_0+t$; we write $q_1$ for this probability. Similarly we also measure the probability that two individuals who are not connected at $t_0$, are neighbours at time $t_0+t$; we denote this probability by $q_0$. In the stationary state (i.e. for large $t_0$), the time, $t$, at which $q_0\approx q_1$ is a good estimate for the time it takes for the network to be `renewed'. Results are shown in Fig.~\ref{fig:LinkProb}. For the parameters used throughout this paper, we estimate the renewal time as $t_{\rm r}\approx6.4$; this is the first time, $t$, for which both $q_0$ and $q_1$ are within $0.1\%$ of their asymptotic value, $q$. Recalling that the flow period is $T=1$, this indicates that the set of neighbours of any one individual in the population is uncorrelated from the set of neighbours of the same individual approximately six and half periods earlier.

\subsection{Conversion into characteristic flow speeds}
The times $t_{\rm q}$ and $t_{\rm r}$ can be used to obtain estimates of the characteristic flow speeds separating the different regimes of dynamics described in the main text. These estimates are obtained by comparing $t_{\rm q}$ and $t_{\rm r}$ to the time-to-fixation for different flow speeds, $S$. 

\begin{figure}[htb!!]
	\center
	\includegraphics[width=0.45\textwidth]{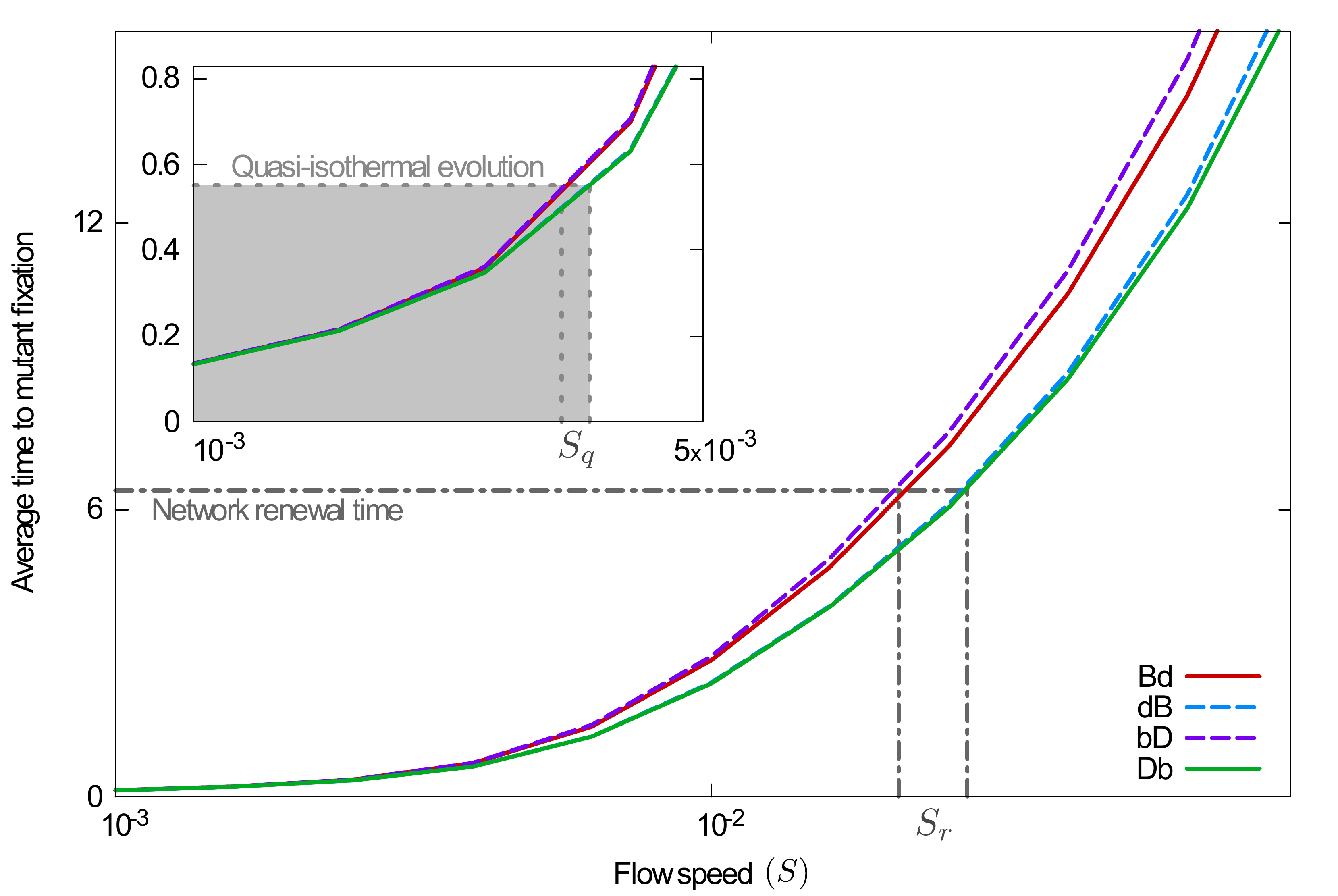}
		\caption{\textbf{Identification of time scales and flow speeds for the different evolutionary regimes.} The time to mutant fixation is plotted for the different evolutionary processes as a function of $S$. The flow speed at which the quasi-isothermal regime ends ($S_q$) is identified as the speed at which the mean time to fixation coincides with the time needed to significantly disrupt the interaction network, obtained in Fig.~\ref{fig:FlowDom}. Similarly, $S_r$ is the flow speed at which the mean fixation time agrees with the network renewal time, obtained in Fig.~\ref{fig:LinkProb}.				}
	\label{fig:S_FixTimesS}
\end{figure}

This is shown in Fig.~\ref{fig:S_FixTimesS}, where we plot the the time, $t_1$, required for a single mutant to reach fixation. Data is shown as a function of $S$.  The network renewal time $t_{\rm r}$ is marked with a dash-dotted line on the vertical axis of the main panel, and $t_{\rm q}$ is marked with a dotted line on the vertical axis of the inset. The flow speeds for which $t_1=t_{\rm q}$ and $t_1=t_{\rm r}$ define flow speeds $S_{\rm q}$ and $S_{\rm r}$, respectively. Due to differences in the mean fixation times, we note that the estimates for $S_{\rm q}$ and $S_{\rm r}$ vary between the different evolutionary update rules.

\begin{figure}[htb!!]
	\center
	\includegraphics[width=0.45\textwidth]{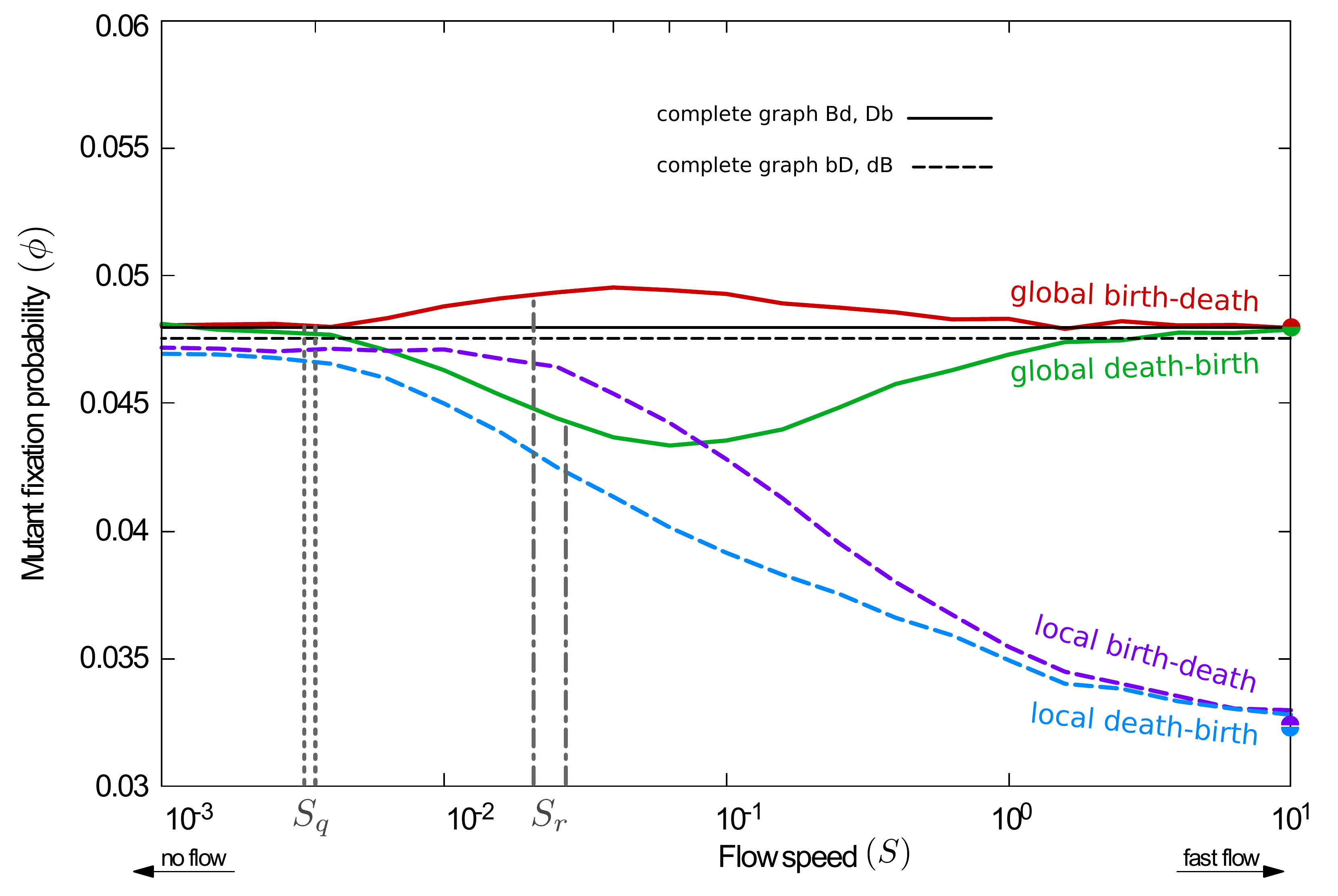}
	\caption{\textbf{Identification of time scales and flow speeds for the different evolutionary regimes.} 
 Fixation probability at different flow speeds for simulations started from a square lattice are shown. $S_q$ roughly corresponds to the speed marking the end of the quasi-isothermal regime; $S_r$ is found to be of the same order of magnitude as the speed at the extrema of fixation probability.
				}
	\label{fig:S_FixTimesS2}
\end{figure}

These flow speeds are shown in the context of the fixation probability at different flow speeds in Fig.~\ref{fig:S_FixTimesS2}. As expected, the  $S_{\rm q}$ (grey dotted lines) marks the end of the quasi-isothermal regime. For flow speeds $S<S_{\rm q}$ the mean time to fixation is shorter than the time $t_q$ it takes the flow to significantly disrupt the initial lattice. This is the quasi-isothermal regime. For $S>S_q$ fixation is usually reached when the lattice has been significantly distorted.

The grey dash-dotted lines in Fig.~\ref{fig:S_FixTimesS2} correspond to $S_{\rm r}$. The location of the extrema of the fixation probability for \Bd ~and \Db ~processes are found at flow speeds of the same order of magnitude as $S_r$. Similar observations were made for the time-to-consensus in a voter model in \cite{Galla2016}.

\pagebreak
\section{Relevance of the number of active links}\label{app:WheelGraphs}
The amplification or suppression of selection, observed when comparing the simulations initialized from connected and unrestricted graphs, can be understood using an argument analogous to the one in ref.~\cite{Antal2006}. The number of active links does not change the probability with which a node is picked in the initial step of an evolutionary event, but it does have an effect on the choice of the second individual.

\begin{figure}[htb!!]
	\center
	\includegraphics[width=0.65\textwidth]{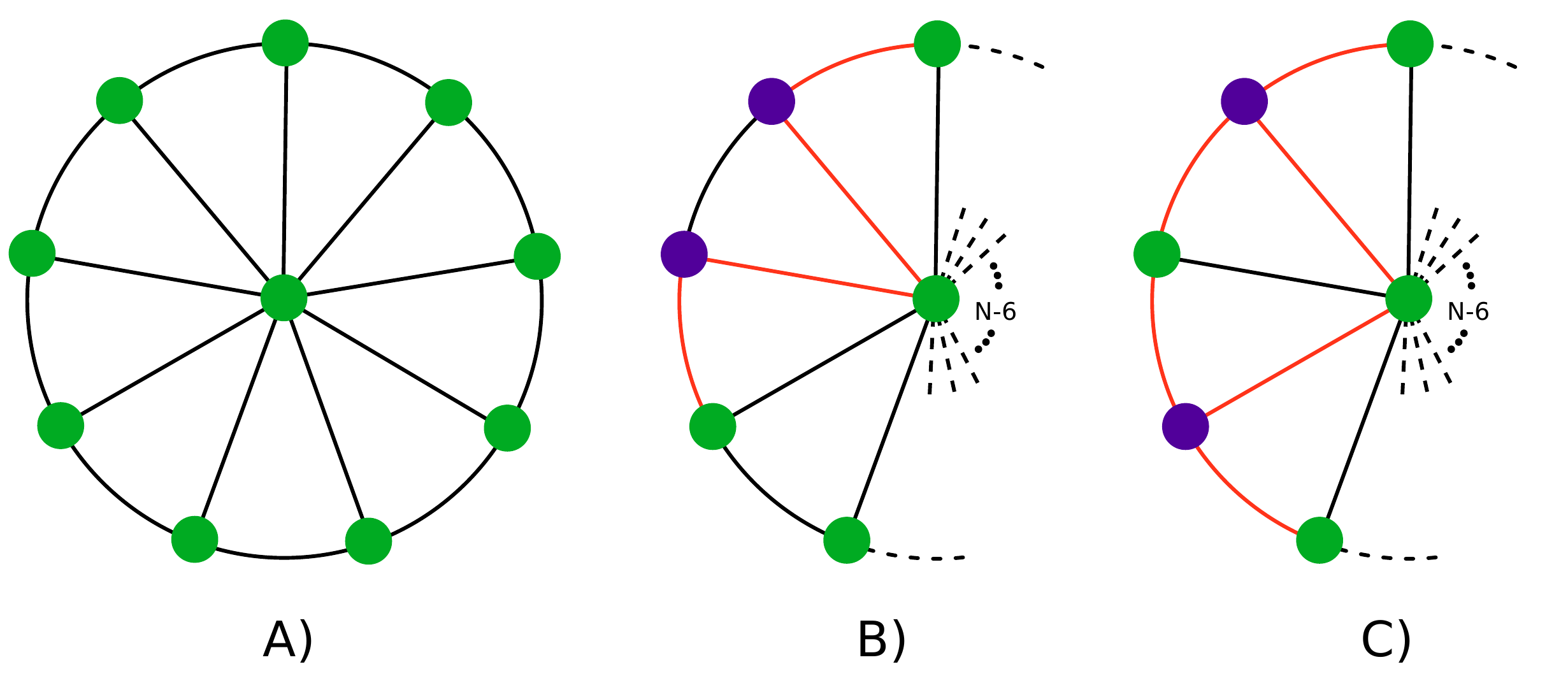}
	\caption{\textbf{Wheel graphs.} 
	A) A sample wheel graph of size 10; B) A portion of a wheel graph with two mutants on adjacent leaves, with $L_{\rm act}=4$; C) A portion of a wheel graph with two mutants on non-adjacent leaves, with $L_{\rm act}=6$. Active links are marked  orange. For B and C, $N-6$ wildtype nodes are not shown.
				}
	\label{fig:Wheels}
\end{figure}

Wheel graphs are convenient to illustrate this. They consist of a central node (hub), connected to $N-1$ nodes organized in a circle around it (leaves), where $N$ is the total size of the graph. The leaves are only connected to the hub and to their nearest neighbours on the rim of the wheel (see Fig.~\ref{fig:Wheels}~A). If two mutants are placed on a graph of this type, their location affects the number of active links ($L_{\rm act}$), even though the total population size, the number of mutants and the degree of either mutants or wildtypes in the population remain unchanged. This is shown in Fig.~\ref{fig:Wheels}~B~and~C.

When the two mutants are neighbours (B), the network has $4$ active links. If they are not neighbours (C), $L_{\rm act}=6$. To be able to have both settings one needs $N\geq5$. In the two cases, both mutants have degree $3$, and there are $N-3$ wildtypes with degree $3$, and one with degree $N-1$. The probability that a mutant is replaced by a wildtype, however, is not the same for the two cases.

To see this, we focus on the case of neutral selection. We write $p_{m , w}$ for the probability that, in a single evolutionary event, a mutant is chosen first and a wildtype second. Similarly $p_{w , m}$ is the probability that a wildtype is chosen first and a mutant second. We then look at the ratio $Q=p_{m , w}/p_{w , m}$, to determine which one of these is more likely.

The probability that the individual picked in the first step is a mutant is $p_m=2/N$; for a wildtype, $p_w=(N-2)/N$. This is the case in both configurations, B and C in Fig.~\ref{fig:Wheels}. For configuration B we then have:
\begin{align*}
	p_{m , w}^B&	= \frac{2}{N}\left[\frac{1}{2}\left(\frac{2}{3}+\frac{2}{3}\right)\right]
							= \frac{4}{3N},
	\\
	p_{w , m}^B&	= \frac{N-2}{N}\left[\frac{1}{N-2}\left(\frac{2}{N-1}+2\frac{1}{3}\right)\right]
							= \frac{2}{N}\frac{N+2}{3(N-1)}.
\end{align*}
From this we find
\begin{equation}
	Q_B = \frac{\frac{4}{3N}}{\frac{2}{N}\frac{N+2}{3(N-1)}}
	     = \frac{2(N-1)}{N+2}
	\label{eq:QB}
\end{equation}

For configuration C we find
\begin{align*}
	p_{m , w}^C&	= \frac{2}{N}\left[\frac{1}{2}\left(\frac{3}{3}+\frac{3}{3}\right)\right]
							= \frac{2}{N},
	\\
	p_{w , m}^C&	= \frac{N-2}{N}\left[\frac{1}{N-2}\left(\frac{2}{N-1}+4\frac{1}{3}\right)\right]
							= \frac{2}{N}\frac{2N+1}{3(N-1)},
\end{align*}
and hence
\begin{equation}
	Q_C = \frac{\frac{2}{N}}{\frac{2}{N}\frac{2N+1}{3(N-1)}}
	     = \frac{3(N-1)}{2N+1}
	\label{eq:QC}
\end{equation}

Comparing Eqs.~(\ref{eq:QB})~and~(\ref{eq:QC}), and assuming $N\geq5$, we find $Q_B>Q_C$, i.e., the scenario with more active links (C) is more likely to result in events in which a wildtype is picked first and a mutant second. For a death-birth process, this means that mutants are more likely to reproduce, and for a birth-death process that they are more likely to die. This is in line with the results obtained in Fig.~\ref{fig:FixProbS_C} of the main text.

A further configuration on a the wheel graphs is possible, placing one of the two mutants in the hub, which results in $L_{\rm act}=N$. A similar analysis leads to $Q=(5N-8)/[N(N-1)]$, which is even smaller than $Q_C$. We note however that the average degree of mutants and wildtypes in this scenario is different from those in the other two settings.

\section{Dual selection processes}\label{app:Dual}
As mentioned in Sec.~\ref{sec:Methods}, the most general process of the birth-death or death-birth type is one that includes selection in both steps. In line with ref.\cite{Herrerias2018}, we call these `dual-selection' processes; we label them \BD~ and \DB, respectively.

\begin{figure}[htb!!]
	\center
	\includegraphics[width=0.61\textwidth]{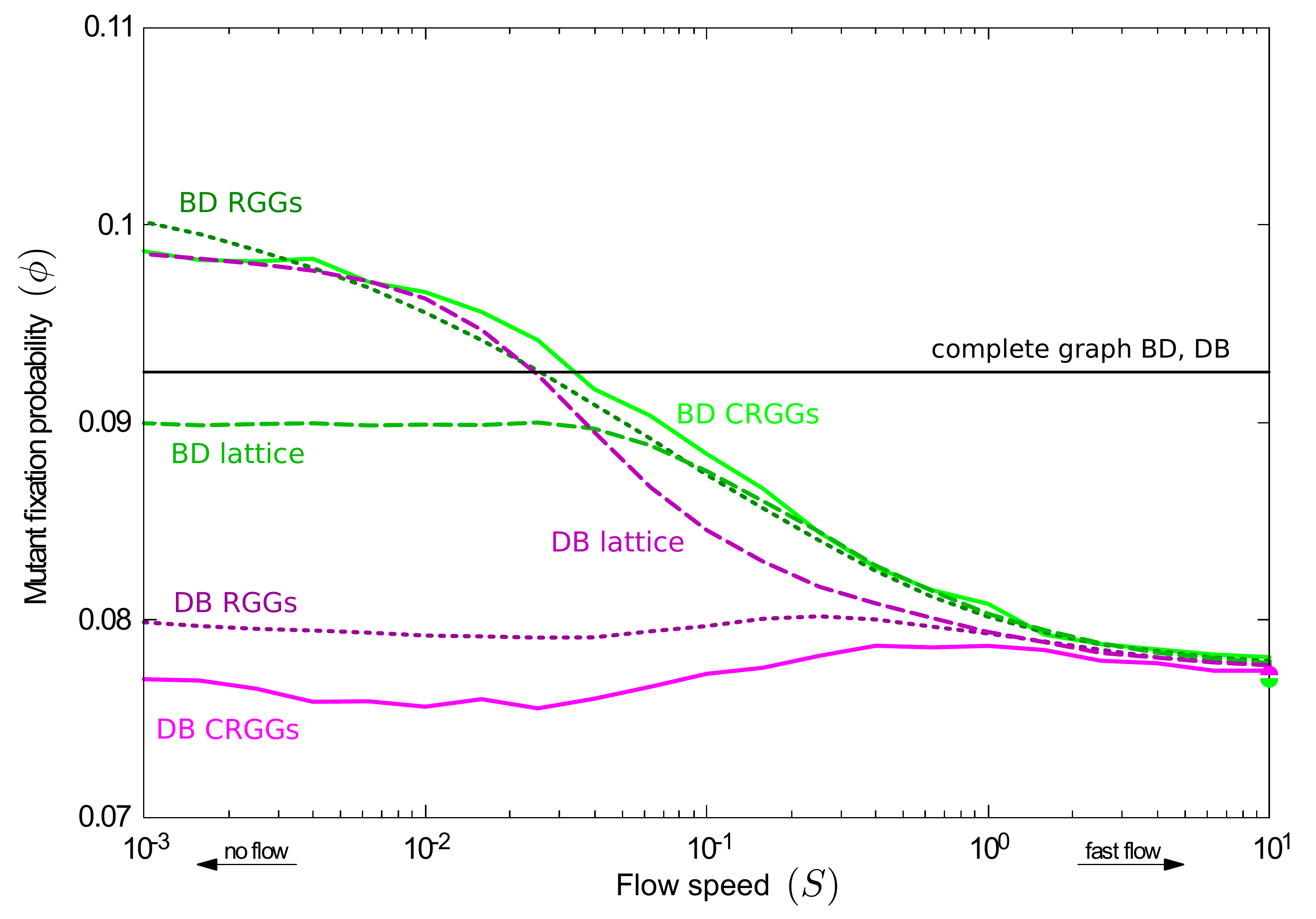}
	\caption{\textbf{Fixation probability at different flow speeds for dual-selection processes.} 
The result for the complete graph is plotted as a reference. Continuous lines correspond to simulations initialized from CRGGs, dotted lines to unrestricted RGGs, and dashed lines to simulations started from a lattice.
				}
	\label{fig:FixProbS_Dual}
\end{figure}

In this section, we present and discuss the simulation results for fixation probabilities at different flow speeds for both dual-selection processes. These are shown in Fig.~\ref{fig:FixProbS_Dual} for all configurations of initial positions studied in the main text: RGGs, CRGGs, and the square lattice.

Naively, one could expect to be able to describe the response of the fixation probabilities to the flow speed in dual-selection processes through a combination of those observed for the corresponding global and local processes. That is, for \BD~ one would expect to see similar features as in \Bd ~and \bD, and the behaviour of \DB~ could be expected to show elements of that of \Db ~and \dB. 
For example, both \BD~ and \DB ~involve local selection, and for fast flows we find that the resulting fixation probability is lower than that of the complete graph, as is the case for both local processes, \bD~ and \dB.

\medskip
For slow flows we make the following observations:
\medskip

{\em Dual-selection birth-death process}: For simulations initialized on RGGs and CRGGs, the fixation probability of the \BD ~process (dotted and continuous green lines in Fig.~\ref{fig:FixProbS_Dual}) is above the one on complete graphs, as is the case for both \Bd ~and \bD ~(see red and purple lines in Fig.~\ref{fig:FixProbS_C}).

If simulations are started from a lattice (dashed green line in Fig.~\ref{fig:FixProbS_Dual}), the fixation probability of the \BD ~process is slightly below $\phi_{CG}$; this could also be expected, as the isothermal theorem does not hold due to the presence of local selection; we note that the fixation probability of the \bD ~process is below that on a complete graph (see purple dashed line in Fig.~\ref{fig:FixProbS_L}).

Interestingly, at slow flow speeds random initial positions act as an amplifier of selection (compared to the complete graph) for the \BD ~process, whereas suppression of selection is observed for lattice initial positions.

\medskip
{\em Dual-selection death-birth process}: The fixation probability of the \DB ~process in simulations started on RGGs and CRGGs (dotted and continuous magenta lines in Fig.~\ref{fig:FixProbS_Dual}) is well below $\phi_{CG}$; we note that the mutant's success for both \Db ~and \dB ~is below that on a complete graph (see green and blue lines in Fig.~\ref{fig:FixProbS_C}), so it is not surprising that the \DB~ process shows this feature as well.

However, if simulations are started from a lattice (dashed magenta line in Fig.~\ref{fig:FixProbS_Dual}), the fixation probability of the \DB ~process is above $\phi_{CG}$. This is different from both the \Db ~or \dB ~processes, who both lead to $\phi\leq\phi_{CG}$ (see green and blue lines in Fig.~\ref{fig:FixProbS_L}). This indicates that simple intuition may fail -- features present in \Db~ and in \dB~ processes may be altered when selection acts in both the death and the birth step. This, we believe is an unexpected observation, which could be pursued in future work. In particular it would be interesting to test when exactly such counter-intuitive behaviour is found when combining local and global selection, i.e., for example, for what types of graphs does this occur, and what common features do these graphs have?

\medskip

We also note that, in contrast to the \BD~process, we find amplification of selection for the \DB ~process when starting from regular lattice and slow flow. For random initial positions we find suppressed selection.

\label{LastPageApp}		



\begin{thebibliography}{77}%
\makeatletter
\providecommand \@ifxundefined [1]{%
 \@ifx{#1\undefined}
}%
\providecommand \@ifnum [1]{%
 \ifnum #1\expandafter \@firstoftwo
 \else \expandafter \@secondoftwo
 \fi
}%
\providecommand \@ifx [1]{%
 \ifx #1\expandafter \@firstoftwo
 \else \expandafter \@secondoftwo
 \fi
}%
\providecommand \natexlab [1]{#1}%
\providecommand \enquote  [1]{``#1''}%
\providecommand \bibnamefont  [1]{#1}%
\providecommand \bibfnamefont [1]{#1}%
\providecommand \citenamefont [1]{#1}%
\providecommand \href@noop [0]{\@secondoftwo}%
\providecommand \href [0]{\begingroup \@sanitize@url \@href}%
\providecommand \@href[1]{\@@startlink{#1}\@@href}%
\providecommand \@@href[1]{\endgroup#1\@@endlink}%
\providecommand \@sanitize@url [0]{\catcode `\\12\catcode `\$12\catcode
  `\&12\catcode `\#12\catcode `\^12\catcode `\_12\catcode `\%12\relax}%
\providecommand \@@startlink[1]{}%
\providecommand \@@endlink[0]{}%
\providecommand \url  [0]{\begingroup\@sanitize@url \@url }%
\providecommand \@url [1]{\endgroup\@href {#1}{\urlprefix }}%
\providecommand \urlprefix  [0]{URL }%
\providecommand \Eprint [0]{\href }%
\providecommand \doibase [0]{http://dx.doi.org/}%
\providecommand \selectlanguage [0]{\@gobble}%
\providecommand \bibinfo  [0]{\@secondoftwo}%
\providecommand \bibfield  [0]{\@secondoftwo}%
\providecommand \translation [1]{[#1]}%
\providecommand \BibitemOpen [0]{}%
\providecommand \bibitemStop [0]{}%
\providecommand \bibitemNoStop [0]{.\EOS\space}%
\providecommand \EOS [0]{\spacefactor3000\relax}%
\providecommand \BibitemShut  [1]{\csname bibitem#1\endcsname}%
\let\auto@bib@innerbib\@empty
\bibitem [{\citenamefont {{Maynard Smith}}(1982)}]{MaynardSmith1982}%
  \BibitemOpen
  \bibfield  {author} {\bibinfo {author} {\bibfnamefont {John}\ \bibnamefont
  {{Maynard Smith}}},\ }\href@noop {} {\emph {\bibinfo {title} {{Evolution and
  the Theory of Games}}}}\ (\bibinfo  {publisher} {Cambridge University
  Press},\ \bibinfo {address} {Cambridge, UK},\ \bibinfo {year}
  {1982})\BibitemShut {NoStop}%
\bibitem [{\citenamefont {Ewens}(2004)}]{Ewens2004}%
  \BibitemOpen
  \bibfield  {author} {\bibinfo {author} {\bibfnamefont {Warren~J.}\
  \bibnamefont {Ewens}},\ }\href@noop {} {\emph {\bibinfo {title}
  {{Mathematical Population Genetics 1: Theoretical Introduction}}}}\ (\bibinfo
   {publisher} {Springer},\ \bibinfo {address} {New York, NY},\ \bibinfo {year}
  {2004})\BibitemShut {NoStop}%
\bibitem [{\citenamefont {Nowak}(2006{\natexlab{a}})}]{Nowak2006a}%
  \BibitemOpen
  \bibfield  {author} {\bibinfo {author} {\bibfnamefont {Martin~A.}\
  \bibnamefont {Nowak}},\ }\href@noop {} {\emph {\bibinfo {title}
  {{Evolutionary Dynamics}}}}\ (\bibinfo  {publisher} {Harvard University
  Press},\ \bibinfo {address} {Cambridge, Massachusetts},\ \bibinfo {year}
  {2006})\BibitemShut {NoStop}%
\bibitem [{\citenamefont {Zukewich}\ \emph {et~al.}(2013)\citenamefont
  {Zukewich}, \citenamefont {Kurella}, \citenamefont {Doebeli},\ and\
  \citenamefont {Hauert}}]{Zukewich2013}%
  \BibitemOpen
  \bibfield  {author} {\bibinfo {author} {\bibfnamefont {Joshua}\ \bibnamefont
  {Zukewich}}, \bibinfo {author} {\bibfnamefont {Venu}\ \bibnamefont
  {Kurella}}, \bibinfo {author} {\bibfnamefont {Michael}\ \bibnamefont
  {Doebeli}}, \ and\ \bibinfo {author} {\bibfnamefont {Christoph}\ \bibnamefont
  {Hauert}},\ }\bibfield  {title} {\enquote {\bibinfo {title} {{Consolidating
  Birth-Death and Death-Birth Processes in Structured Populations}},}\ }\href
  {\doibase 10.1371/journal.pone.0054639} {\bibfield  {journal} {\bibinfo
  {journal} {PLoS ONE}\ }\textbf {\bibinfo {volume} {8}},\ \bibinfo {pages}
  {e54639} (\bibinfo {year} {2013})}\BibitemShut {NoStop}%
\bibitem [{\citenamefont {Kaveh}\ \emph {et~al.}(2015)\citenamefont {Kaveh},
  \citenamefont {Komarova},\ and\ \citenamefont {Kohandel}}]{Kaveh2015}%
  \BibitemOpen
  \bibfield  {author} {\bibinfo {author} {\bibfnamefont {Kamran}\ \bibnamefont
  {Kaveh}}, \bibinfo {author} {\bibfnamefont {Natalia~L.}\ \bibnamefont
  {Komarova}}, \ and\ \bibinfo {author} {\bibfnamefont {Mohammad}\ \bibnamefont
  {Kohandel}},\ }\bibfield  {title} {\enquote {\bibinfo {title} {{The duality
  of spatial death-birth and birth-death processes and limitations of the
  isothermal theorem}},}\ }\href {\doibase 10.1098/rsos.140465} {\bibfield
  {journal} {\bibinfo  {journal} {Royal Society Open Science}\ }\textbf
  {\bibinfo {volume} {2}},\ \bibinfo {pages} {140465--140465} (\bibinfo {year}
  {2015})}\BibitemShut {NoStop}%
\bibitem [{\citenamefont {Hindersin}\ and\ \citenamefont
  {Traulsen}(2015)}]{Hindersin2015}%
  \BibitemOpen
  \bibfield  {author} {\bibinfo {author} {\bibfnamefont {Laura}\ \bibnamefont
  {Hindersin}}\ and\ \bibinfo {author} {\bibfnamefont {Arne}\ \bibnamefont
  {Traulsen}},\ }\bibfield  {title} {\enquote {\bibinfo {title} {{Most
  undirected random graphs are amplifiers of selection for Birth-death
  dynamics, but suppressors of selection for death-Birth dynamics}},}\ }\href
  {\doibase 10.1371/journal.pcbi.1004437} {\bibfield  {journal} {\bibinfo
  {journal} {PLOS Computational Biology}\ }\textbf {\bibinfo {volume} {11}},\
  \bibinfo {pages} {e1004437} (\bibinfo {year} {2015})}\BibitemShut {NoStop}%
\bibitem [{\citenamefont {Kerr}\ \emph {et~al.}(2002)\citenamefont {Kerr},
  \citenamefont {Riley}, \citenamefont {Feldman},\ and\ \citenamefont
  {Bohannan}}]{Kerr2002}%
  \BibitemOpen
  \bibfield  {author} {\bibinfo {author} {\bibfnamefont {Benjamin}\
  \bibnamefont {Kerr}}, \bibinfo {author} {\bibfnamefont {Margaret~A.}\
  \bibnamefont {Riley}}, \bibinfo {author} {\bibfnamefont {Marcus~W.}\
  \bibnamefont {Feldman}}, \ and\ \bibinfo {author} {\bibfnamefont {Brendan
  J.~M.}\ \bibnamefont {Bohannan}},\ }\bibfield  {title} {\enquote {\bibinfo
  {title} {{Local dispersal promotes biodiversity in a real-life game of
  rock-paper-scissors}},}\ }\href {\doibase 10.1038/nature00823} {\bibfield
  {journal} {\bibinfo  {journal} {Nature}\ }\textbf {\bibinfo {volume} {418}},\
  \bibinfo {pages} {171--174} (\bibinfo {year} {2002})}\BibitemShut {NoStop}%
\bibitem [{\citenamefont {Habets}\ \emph {et~al.}(2006)\citenamefont {Habets},
  \citenamefont {Rozen}, \citenamefont {Hoekstra},\ and\ \citenamefont
  {de~Visser}}]{Habets2006}%
  \BibitemOpen
  \bibfield  {author} {\bibinfo {author} {\bibfnamefont {Michelle G J~L}\
  \bibnamefont {Habets}}, \bibinfo {author} {\bibfnamefont {Daniel~E.}\
  \bibnamefont {Rozen}}, \bibinfo {author} {\bibfnamefont {Rolf~F.}\
  \bibnamefont {Hoekstra}}, \ and\ \bibinfo {author} {\bibfnamefont {J.~Arjan
  G.~M.}\ \bibnamefont {de~Visser}},\ }\bibfield  {title} {\enquote {\bibinfo
  {title} {{The effect of population structure on the adaptive radiation of
  microbial populations evolving in spatially structured environments}},}\
  }\href {\doibase 10.1111/j.1461-0248.2006.00955.x} {\bibfield  {journal}
  {\bibinfo  {journal} {Ecology Letters}\ }\textbf {\bibinfo {volume} {9}},\
  \bibinfo {pages} {1041--1048} (\bibinfo {year} {2006})}\BibitemShut {NoStop}%
\bibitem [{\citenamefont {Perfeito}\ \emph {et~al.}(2008)\citenamefont
  {Perfeito}, \citenamefont {Pereira}, \citenamefont {Campos},\ and\
  \citenamefont {Gordo}}]{Perfeito2008a}%
  \BibitemOpen
  \bibfield  {author} {\bibinfo {author} {\bibfnamefont {Lilia}\ \bibnamefont
  {Perfeito}}, \bibinfo {author} {\bibfnamefont {M.~In{\^{e}}s}\ \bibnamefont
  {Pereira}}, \bibinfo {author} {\bibfnamefont {Paulo~R.A.}\ \bibnamefont
  {Campos}}, \ and\ \bibinfo {author} {\bibfnamefont {Isabel}\ \bibnamefont
  {Gordo}},\ }\bibfield  {title} {\enquote {\bibinfo {title} {{The effect of
  spatial structure on adaptation in Escherichia coli}},}\ }\href {\doibase
  10.1098/rsbl.2007.0481} {\bibfield  {journal} {\bibinfo  {journal} {Biology
  Letters}\ }\textbf {\bibinfo {volume} {4}},\ \bibinfo {pages} {57--59}
  (\bibinfo {year} {2008})}\BibitemShut {NoStop}%
\bibitem [{\citenamefont {Nowak}\ and\ \citenamefont {May}(1992)}]{Nowak1992}%
  \BibitemOpen
  \bibfield  {author} {\bibinfo {author} {\bibfnamefont {Martin~A.}\
  \bibnamefont {Nowak}}\ and\ \bibinfo {author} {\bibfnamefont {Robert~M.}\
  \bibnamefont {May}},\ }\bibfield  {title} {\enquote {\bibinfo {title}
  {{Evolutionary games and spatial chaos}},}\ }\href {\doibase
  10.1038/359826a0} {\bibfield  {journal} {\bibinfo  {journal} {Nature}\
  }\textbf {\bibinfo {volume} {359}},\ \bibinfo {pages} {826--829} (\bibinfo
  {year} {1992})}\BibitemShut {NoStop}%
\bibitem [{\citenamefont {Ebel}\ and\ \citenamefont
  {Bornholdt}(2002)}]{Ebel2002}%
  \BibitemOpen
  \bibfield  {author} {\bibinfo {author} {\bibfnamefont {Holger}\ \bibnamefont
  {Ebel}}\ and\ \bibinfo {author} {\bibfnamefont {Stefan}\ \bibnamefont
  {Bornholdt}},\ }\bibfield  {title} {\enquote {\bibinfo {title}
  {{Coevolutionary games on networks}},}\ }\href {\doibase
  10.1103/PhysRevE.66.056118} {\bibfield  {journal} {\bibinfo  {journal}
  {Physical Review E}\ }\textbf {\bibinfo {volume} {66}},\ \bibinfo {pages}
  {056118} (\bibinfo {year} {2002})}\BibitemShut {NoStop}%
\bibitem [{\citenamefont {Nowak}(2006{\natexlab{b}})}]{Nowak2006}%
  \BibitemOpen
  \bibfield  {author} {\bibinfo {author} {\bibfnamefont {Martin~A.}\
  \bibnamefont {Nowak}},\ }\bibfield  {title} {\enquote {\bibinfo {title}
  {{Five rules for the evolution of cooperation}},}\ }\href {\doibase
  10.1126/science.1133755} {\bibfield  {journal} {\bibinfo  {journal}
  {Science}\ }\textbf {\bibinfo {volume} {314}},\ \bibinfo {pages} {1560--1563}
  (\bibinfo {year} {2006}{\natexlab{b}})}\BibitemShut {NoStop}%
\bibitem [{\citenamefont {Santos}\ \emph {et~al.}(2006)\citenamefont {Santos},
  \citenamefont {Pacheco},\ and\ \citenamefont {Lenaerts}}]{Santos2006}%
  \BibitemOpen
  \bibfield  {author} {\bibinfo {author} {\bibfnamefont {Francisco~C.}\
  \bibnamefont {Santos}}, \bibinfo {author} {\bibfnamefont {Jorge~M.}\
  \bibnamefont {Pacheco}}, \ and\ \bibinfo {author} {\bibfnamefont {Tom}\
  \bibnamefont {Lenaerts}},\ }\bibfield  {title} {\enquote {\bibinfo {title}
  {{Evolutionary dynamics of social dilemmas in structured heterogeneous
  populations}},}\ }\href {\doibase 10.1073/pnas.0508201103} {\bibfield
  {journal} {\bibinfo  {journal} {Proceedings of the National Academy of
  Sciences}\ }\textbf {\bibinfo {volume} {103}},\ \bibinfo {pages} {3490--3494}
  (\bibinfo {year} {2006})}\BibitemShut {NoStop}%
\bibitem [{\citenamefont {Ohtsuki}\ \emph {et~al.}(2006)\citenamefont
  {Ohtsuki}, \citenamefont {Hauert}, \citenamefont {Lieberman},\ and\
  \citenamefont {Nowak}}]{Ohtsuki2006}%
  \BibitemOpen
  \bibfield  {author} {\bibinfo {author} {\bibfnamefont {Hisashi}\ \bibnamefont
  {Ohtsuki}}, \bibinfo {author} {\bibfnamefont {Christoph}\ \bibnamefont
  {Hauert}}, \bibinfo {author} {\bibfnamefont {Erez}\ \bibnamefont
  {Lieberman}}, \ and\ \bibinfo {author} {\bibfnamefont {Martin~A.}\
  \bibnamefont {Nowak}},\ }\bibfield  {title} {\enquote {\bibinfo {title} {{A
  simple rule for the evolution of cooperation on graphs and social
  networks}},}\ }\href {\doibase 10.1038/nature04605} {\bibfield  {journal}
  {\bibinfo  {journal} {Nature}\ }\textbf {\bibinfo {volume} {441}},\ \bibinfo
  {pages} {502--505} (\bibinfo {year} {2006})}\BibitemShut {NoStop}%
\bibitem [{\citenamefont {Nowak}\ \emph {et~al.}(2010)\citenamefont {Nowak},
  \citenamefont {Tarnita},\ and\ \citenamefont {Antal}}]{Nowak2010}%
  \BibitemOpen
  \bibfield  {author} {\bibinfo {author} {\bibfnamefont {Martin~A.}\
  \bibnamefont {Nowak}}, \bibinfo {author} {\bibfnamefont {Corina~E.}\
  \bibnamefont {Tarnita}}, \ and\ \bibinfo {author} {\bibfnamefont {Tibor}\
  \bibnamefont {Antal}},\ }\bibfield  {title} {\enquote {\bibinfo {title}
  {{Evolutionary dynamics in structured populations}},}\ }\href {\doibase
  10.1098/rstb.2009.0215} {\bibfield  {journal} {\bibinfo  {journal}
  {Philosophical Transactions of the Royal Society B: Biological Sciences}\
  }\textbf {\bibinfo {volume} {365}},\ \bibinfo {pages} {19--30} (\bibinfo
  {year} {2010})}\BibitemShut {NoStop}%
\bibitem [{\citenamefont {Szab{\'{o}}}\ and\ \citenamefont
  {F{\'{a}}th}(2007)}]{Szabo2007}%
  \BibitemOpen
  \bibfield  {author} {\bibinfo {author} {\bibfnamefont {Gy{\"{o}}rgy}\
  \bibnamefont {Szab{\'{o}}}}\ and\ \bibinfo {author} {\bibfnamefont
  {G{\'{a}}bor}\ \bibnamefont {F{\'{a}}th}},\ }\bibfield  {title} {\enquote
  {\bibinfo {title} {{Evolutionary games on graphs}},}\ }\href {\doibase
  10.1016/j.physrep.2007.04.004} {\bibfield  {journal} {\bibinfo  {journal}
  {Physics Reports}\ }\textbf {\bibinfo {volume} {446}},\ \bibinfo {pages}
  {97--216} (\bibinfo {year} {2007})}\BibitemShut {NoStop}%
\bibitem [{\citenamefont {Gross}\ and\ \citenamefont
  {Blasius}(2008)}]{Gross2008}%
  \BibitemOpen
  \bibfield  {author} {\bibinfo {author} {\bibfnamefont {Thilo}\ \bibnamefont
  {Gross}}\ and\ \bibinfo {author} {\bibfnamefont {Bernd}\ \bibnamefont
  {Blasius}},\ }\bibfield  {title} {\enquote {\bibinfo {title} {{Adaptive
  coevolutionary networks: A review}},}\ }\href {\doibase
  10.1098/rsif.2007.1229} {\bibfield  {journal} {\bibinfo  {journal} {Journal
  of the Royal Society Interface}\ }\textbf {\bibinfo {volume} {5}},\ \bibinfo
  {pages} {259--271} (\bibinfo {year} {2008})}  \BibitemShut {NoStop}%
\bibitem [{\citenamefont {Poncela}\ \emph {et~al.}(2009)\citenamefont
  {Poncela}, \citenamefont {G{\'{o}}mez-Garde{\~{n}}es}, \citenamefont
  {Traulsen},\ and\ \citenamefont {Moreno}}]{Poncela2009}%
  \BibitemOpen
  \bibfield  {author} {\bibinfo {author} {\bibfnamefont {Julia}\ \bibnamefont
  {Poncela}}, \bibinfo {author} {\bibfnamefont {Jes{\'{u}}s}\ \bibnamefont
  {G{\'{o}}mez-Garde{\~{n}}es}}, \bibinfo {author} {\bibfnamefont {Arne}\
  \bibnamefont {Traulsen}}, \ and\ \bibinfo {author} {\bibfnamefont {Yamir}\
  \bibnamefont {Moreno}},\ }\bibfield  {title} {\enquote {\bibinfo {title}
  {{Evolutionary game dynamics in a growing structured population}},}\ }\href
  {\doibase 10.1088/1367-2630/11/8/083031} {\bibfield  {journal} {\bibinfo
  {journal} {New Journal of Physics}\ }\textbf {\bibinfo {volume} {11}},\
  \bibinfo {pages} {083031} (\bibinfo {year} {2009})}\BibitemShut {NoStop}%
\bibitem [{\citenamefont {Masuda}\ and\ \citenamefont
  {Ohtsuki}(2009)}]{Masuda2009a}%
  \BibitemOpen
  \bibfield  {author} {\bibinfo {author} {\bibfnamefont {Naoki}\ \bibnamefont
  {Masuda}}\ and\ \bibinfo {author} {\bibfnamefont {Hisashi}\ \bibnamefont
  {Ohtsuki}},\ }\bibfield  {title} {\enquote {\bibinfo {title} {{Evolutionary
  dynamics and fixation probabilities in directed networks}},}\ }\href
  {\doibase 10.1088/1367-2630/11/3/033012} {\bibfield  {journal} {\bibinfo
  {journal} {New Journal of Physics}\ }\textbf {\bibinfo {volume} {11}},\
  \bibinfo {pages} {033012} (\bibinfo {year} {2009})}\BibitemShut {NoStop}%
\bibitem [{\citenamefont {Broom}\ \emph {et~al.}(2011)\citenamefont {Broom},
  \citenamefont {Rycht{\'{a}}{\v{r}}},\ and\ \citenamefont
  {Stadler}}]{Broom2011}%
  \BibitemOpen
  \bibfield  {author} {\bibinfo {author} {\bibfnamefont {Mark}\ \bibnamefont
  {Broom}}, \bibinfo {author} {\bibfnamefont {Jan}\ \bibnamefont
  {Rycht{\'{a}}{\v{r}}}}, \ and\ \bibinfo {author} {\bibfnamefont {B.~T.}\
  \bibnamefont {Stadler}},\ }\bibfield  {title} {\enquote {\bibinfo {title}
  {{Evolutionary Dynamics on Graphs - the Effect of Graph Structure and Initial
  Placement on Mutant Spread}},}\ }\href {\doibase
  10.1080/15598608.2011.10412035} {\bibfield  {journal} {\bibinfo  {journal}
  {Journal of Statistical Theory and Practice}\ }\textbf {\bibinfo {volume}
  {5}},\ \bibinfo {pages} {369--381} (\bibinfo {year} {2011})}\BibitemShut
  {NoStop}%
\bibitem [{\citenamefont {Perc}\ and\ \citenamefont
  {Szolnoki}(2012)}]{Perc2012}%
  \BibitemOpen
  \bibfield  {author} {\bibinfo {author} {\bibfnamefont {Matja{\v{z}}}\
  \bibnamefont {Perc}}\ and\ \bibinfo {author} {\bibfnamefont {Attila}\
  \bibnamefont {Szolnoki}},\ }\bibfield  {title} {\enquote {\bibinfo {title}
  {{Self-organization of punishment in structured populations}},}\ }\href
  {\doibase 10.1088/1367-2630/14/4/043013} {\bibfield  {journal} {\bibinfo
  {journal} {New Journal of Physics}\ }\textbf {\bibinfo {volume} {14}},\
  \bibinfo {pages} {043013} (\bibinfo {year} {2012})}\BibitemShut {NoStop}%
\bibitem [{\citenamefont {Wang}\ \emph {et~al.}(2013)\citenamefont {Wang},
  \citenamefont {Szolnoki},\ and\ \citenamefont {Perc}}]{Wang2013}%
  \BibitemOpen
  \bibfield  {author} {\bibinfo {author} {\bibfnamefont {Zhen}\ \bibnamefont
  {Wang}}, \bibinfo {author} {\bibfnamefont {Attila}\ \bibnamefont {Szolnoki}},
  \ and\ \bibinfo {author} {\bibfnamefont {Matja{\v{z}}}\ \bibnamefont
  {Perc}},\ }\bibfield  {title} {\enquote {\bibinfo {title} {{Interdependent
  network reciprocity in evolutionary games}},}\ }\href {\doibase
  10.1038/srep01183} {\bibfield  {journal} {\bibinfo  {journal} {Scientific
  Reports}\ }\textbf {\bibinfo {volume} {3}},\ \bibinfo {pages} {1183}
  (\bibinfo {year} {2013})}\BibitemShut {NoStop}%
\bibitem [{\citenamefont {Perc}\ \emph {et~al.}(2013)\citenamefont {Perc},
  \citenamefont {Gomez-Gardenes}, \citenamefont {Szolnoki}, \citenamefont
  {Floria},\ and\ \citenamefont {Moreno}}]{Perc2013}%
  \BibitemOpen
  \bibfield  {author} {\bibinfo {author} {\bibfnamefont {Matja{\v{z}}}\
  \bibnamefont {Perc}}, \bibinfo {author} {\bibfnamefont {J.}~\bibnamefont
  {Gomez-Gardenes}}, \bibinfo {author} {\bibfnamefont {Attila}\ \bibnamefont
  {Szolnoki}}, \bibinfo {author} {\bibfnamefont {L.~M.}\ \bibnamefont
  {Floria}}, \ and\ \bibinfo {author} {\bibfnamefont {Yamir}\ \bibnamefont
  {Moreno}},\ }\bibfield  {title} {\enquote {\bibinfo {title} {{Evolutionary
  dynamics of group interactions on structured populations: a review}},}\
  }\href {\doibase 10.1098/rsif.2012.0997} {\bibfield  {journal} {\bibinfo
  {journal} {Journal of The Royal Society Interface}\ }\textbf {\bibinfo
  {volume} {10}},\ \bibinfo {pages} {20120997--20120997} (\bibinfo {year}
  {2013})}\BibitemShut {NoStop}%
\bibitem [{\citenamefont {Hindersin}\ and\ \citenamefont
  {Traulsen}(2014)}]{Hindersin2014}%
  \BibitemOpen
  \bibfield  {author} {\bibinfo {author} {\bibfnamefont {Laura}\ \bibnamefont
  {Hindersin}}\ and\ \bibinfo {author} {\bibfnamefont {Arne}\ \bibnamefont
  {Traulsen}},\ }\bibfield  {title} {\enquote {\bibinfo {title}
  {{Counterintuitive properties of the fixation time in network-structured
  populations}},}\ }\href {\doibase 10.1098/rsif.2014.0606} {\bibfield
  {journal} {\bibinfo  {journal} {Journal of The Royal Society Interface}\
  }\textbf {\bibinfo {volume} {11}},\ \bibinfo {pages} {20140606--20140606}
  (\bibinfo {year} {2014})}\BibitemShut {NoStop}%
\bibitem [{\citenamefont {Jiang}\ \emph {et~al.}(2014)\citenamefont {Jiang},
  \citenamefont {Chen},\ and\ \citenamefont {Liu}}]{Jiang2014}%
  \BibitemOpen
  \bibfield  {author} {\bibinfo {author} {\bibfnamefont {Chunxiao}\
  \bibnamefont {Jiang}}, \bibinfo {author} {\bibfnamefont {Yan}\ \bibnamefont
  {Chen}}, \ and\ \bibinfo {author} {\bibfnamefont {K.~J.~Ray}\ \bibnamefont
  {Liu}},\ }\bibfield  {title} {\enquote {\bibinfo {title} {{Evolutionary
  Dynamics of Information Diffusion Over Social Networks}},}\ }\href {\doibase
  10.1109/TSP.2014.2339799} {\bibfield  {journal} {\bibinfo  {journal} {IEEE
  Transactions on Signal Processing}\ }\textbf {\bibinfo {volume} {62}},\
  \bibinfo {pages} {4573--4586} (\bibinfo {year} {2014})}\BibitemShut {NoStop}%
\bibitem [{\citenamefont {Allen}\ \emph {et~al.}(2017)\citenamefont {Allen},
  \citenamefont {Lippner}, \citenamefont {Chen}, \citenamefont {Fotouhi},
  \citenamefont {Momeni}, \citenamefont {Yau},\ and\ \citenamefont
  {Nowak}}]{Allen2017}%
  \BibitemOpen
  \bibfield  {author} {\bibinfo {author} {\bibfnamefont {Benjamin}\
  \bibnamefont {Allen}}, \bibinfo {author} {\bibfnamefont {Gabor}\ \bibnamefont
  {Lippner}}, \bibinfo {author} {\bibfnamefont {Yu-Ting}\ \bibnamefont {Chen}},
  \bibinfo {author} {\bibfnamefont {Babak}\ \bibnamefont {Fotouhi}}, \bibinfo
  {author} {\bibfnamefont {Naghmeh}\ \bibnamefont {Momeni}}, \bibinfo {author}
  {\bibfnamefont {Shing-Tung}\ \bibnamefont {Yau}}, \ and\ \bibinfo {author}
  {\bibfnamefont {Martin~A.}\ \bibnamefont {Nowak}},\ }\bibfield  {title}
  {\enquote {\bibinfo {title} {{Evolutionary dynamics on any population
  structure}},}\ }\href {\doibase 10.1038/nature21723} {\bibfield  {journal}
  {\bibinfo  {journal} {Nature}\ }\textbf {\bibinfo {volume} {544}},\ \bibinfo
  {pages} {227--230} (\bibinfo {year} {2017})}\BibitemShut {NoStop}%
\bibitem [{\citenamefont {Pavlogiannis}\ \emph {et~al.}(2018)\citenamefont
  {Pavlogiannis}, \citenamefont {Tkadlec}, \citenamefont {Chatterjee},\ and\
  \citenamefont {Nowak}}]{Pavlogiannisa2018}%
  \BibitemOpen
  \bibfield  {author} {\bibinfo {author} {\bibfnamefont {Andreas}\ \bibnamefont
  {Pavlogiannis}}, \bibinfo {author} {\bibfnamefont {Josef}\ \bibnamefont
  {Tkadlec}}, \bibinfo {author} {\bibfnamefont {Krishnendu}\ \bibnamefont
  {Chatterjee}}, \ and\ \bibinfo {author} {\bibfnamefont {Martin~A.}\
  \bibnamefont {Nowak}},\ }\bibfield  {title} {\enquote {\bibinfo {title}
  {{Construction of arbitrarily strong amplifiers of natural selection using
  evolutionary graph theory}},}\ }\href {\doibase 10.1038/s42003-018-0078-7}
  {\bibfield  {journal} {\bibinfo  {journal} {Communications Biology}\ }\textbf
  {\bibinfo {volume} {1}},\ \bibinfo {pages} {71} (\bibinfo {year}
  {2018})}\BibitemShut {NoStop}%
\bibitem [{\citenamefont {Lieberman}\ \emph {et~al.}(2005)\citenamefont
  {Lieberman}, \citenamefont {Hauert},\ and\ \citenamefont
  {Nowak}}]{Lieberman2005}%
  \BibitemOpen
  \bibfield  {author} {\bibinfo {author} {\bibfnamefont {Erez}\ \bibnamefont
  {Lieberman}}, \bibinfo {author} {\bibfnamefont {Christoph}\ \bibnamefont
  {Hauert}}, \ and\ \bibinfo {author} {\bibfnamefont {Martin~A.}\ \bibnamefont
  {Nowak}},\ }\bibfield  {title} {\enquote {\bibinfo {title} {{Evolutionary
  dynamics on graphs}},}\ }\href {\doibase 10.1038/nature03204} {\bibfield
  {journal} {\bibinfo  {journal} {Nature}\ }\textbf {\bibinfo {volume} {433}},\
  \bibinfo {pages} {312--316} (\bibinfo {year} {2005})}\BibitemShut {NoStop}%
\bibitem [{\citenamefont {Shakarian}\ \emph {et~al.}(2012)\citenamefont
  {Shakarian}, \citenamefont {Roos},\ and\ \citenamefont
  {Johnson}}]{Shakarian2012}%
  \BibitemOpen
  \bibfield  {author} {\bibinfo {author} {\bibfnamefont {Paulo}\ \bibnamefont
  {Shakarian}}, \bibinfo {author} {\bibfnamefont {Patrick}\ \bibnamefont
  {Roos}}, \ and\ \bibinfo {author} {\bibfnamefont {Anthony}\ \bibnamefont
  {Johnson}},\ }\bibfield  {title} {\enquote {\bibinfo {title} {{A review of
  evolutionary graph theory with applications to game theory}},}\ }\href
  {\doibase 10.1016/j.biosystems.2011.09.006} {\bibfield  {journal} {\bibinfo
  {journal} {BioSystems}\ }\textbf {\bibinfo {volume} {107}},\ \bibinfo {pages}
  {66--80} (\bibinfo {year} {2012})}\BibitemShut {NoStop}%
\bibitem [{\citenamefont {Broom}\ and\ \citenamefont
  {Rycht{\'{a}}{\v{r}}}(2008)}]{Broom2008}%
  \BibitemOpen
  \bibfield  {author} {\bibinfo {author} {\bibfnamefont {Mark}\ \bibnamefont
  {Broom}}\ and\ \bibinfo {author} {\bibfnamefont {Jan}\ \bibnamefont
  {Rycht{\'{a}}{\v{r}}}},\ }\bibfield  {title} {\enquote {\bibinfo {title} {{An
  analysis of the fixation probability of a mutant on special classes of
  non-directed graphs}},}\ }\href {\doibase 10.1098/rspa.2008.0058} {\bibfield
  {journal} {\bibinfo  {journal} {Proceedings of the Royal Society A:
  Mathematical, Physical and Engineering Sciences}\ }\textbf {\bibinfo {volume}
  {464}},\ \bibinfo {pages} {2609--2627} (\bibinfo {year} {2008})}\BibitemShut
  {NoStop}%
\bibitem [{\citenamefont {Broom}\ \emph {et~al.}(2009)\citenamefont {Broom},
  \citenamefont {Rycht{\'{a}}{\v{r}}},\ and\ \citenamefont
  {Stadler}}]{Broom2009}%
  \BibitemOpen
  \bibfield  {author} {\bibinfo {author} {\bibfnamefont {Mark}\ \bibnamefont
  {Broom}}, \bibinfo {author} {\bibfnamefont {Jan}\ \bibnamefont
  {Rycht{\'{a}}{\v{r}}}}, \ and\ \bibinfo {author} {\bibfnamefont
  {B.}~\bibnamefont {Stadler}},\ }\bibfield  {title} {\enquote {\bibinfo
  {title} {{Evolutionary dynamics on small-order graphs}},}\ }\href {\doibase
  10.1080/09720502.2009.10700618} {\bibfield  {journal} {\bibinfo  {journal}
  {Journal of Interdisciplinary Mathematics}\ }\textbf {\bibinfo {volume}
  {12}},\ \bibinfo {pages} {129--140} (\bibinfo {year} {2009})}\BibitemShut
  {NoStop}%
\bibitem [{\citenamefont {Houchmandzadeh}\ and\ \citenamefont
  {Vallade}(2011{\natexlab{a}})}]{Houchmandzadeh2011a}%
  \BibitemOpen
  \bibfield  {author} {\bibinfo {author} {\bibfnamefont {Bahram}\ \bibnamefont
  {Houchmandzadeh}}\ and\ \bibinfo {author} {\bibfnamefont {Marcel}\
  \bibnamefont {Vallade}},\ }\bibfield  {title} {\enquote {\bibinfo {title}
  {{The fixation probability of a beneficial mutation in a geographically
  structured population}},}\ }\href {\doibase 10.1088/1367-2630/13/7/073020}
  {\bibfield  {journal} {\bibinfo  {journal} {New Journal of Physics}\ }\textbf
  {\bibinfo {volume} {13}},\ \bibinfo {pages} {073020} (\bibinfo {year}
  {2011}{\natexlab{a}})}\BibitemShut {NoStop}%
\bibitem [{\citenamefont {Pattni}\ \emph {et~al.}(2015)\citenamefont {Pattni},
  \citenamefont {Broom}, \citenamefont {Rycht{\'{a}}{\v{r}}},\ and\
  \citenamefont {Silvers}}]{Pattni2015}%
  \BibitemOpen
  \bibfield  {author} {\bibinfo {author} {\bibfnamefont {Karan}\ \bibnamefont
  {Pattni}}, \bibinfo {author} {\bibfnamefont {Mark}\ \bibnamefont {Broom}},
  \bibinfo {author} {\bibfnamefont {Jan}\ \bibnamefont {Rycht{\'{a}}{\v{r}}}},
  \ and\ \bibinfo {author} {\bibfnamefont {Lara~J.}\ \bibnamefont {Silvers}},\
  }\bibfield  {title} {\enquote {\bibinfo {title} {{Evolutionary graph theory
  revisited: when is an evolutionary process equivalent to the Moran
  process?}}}\ }\href {\doibase 10.1098/rspa.2015.0334} {\bibfield  {journal}
  {\bibinfo  {journal} {Proceedings of the Royal Society A: Mathematical,
  Physical and Engineering Science}\ }\textbf {\bibinfo {volume} {471}},\
  \bibinfo {pages} {20150334} (\bibinfo {year} {2015})}\BibitemShut {NoStop}%
\bibitem [{\citenamefont {Lushi}\ \emph {et~al.}(2014)\citenamefont {Lushi},
  \citenamefont {Wioland},\ and\ \citenamefont {Goldstein}}]{Lushi2014}%
  \BibitemOpen
  \bibfield  {author} {\bibinfo {author} {\bibfnamefont {Enkeleida}\
  \bibnamefont {Lushi}}, \bibinfo {author} {\bibfnamefont {Hugo}\ \bibnamefont
  {Wioland}}, \ and\ \bibinfo {author} {\bibfnamefont {Raymond~E.}\
  \bibnamefont {Goldstein}},\ }\bibfield  {title} {\enquote {\bibinfo {title}
  {{Fluid flows created by swimming bacteria drive self-organization in
  confined suspensions}},}\ }\href {\doibase 10.1073/pnas.1405698111}
  {\bibfield  {journal} {\bibinfo  {journal} {Proceedings of the National
  Academy of Sciences}\ }\textbf {\bibinfo {volume} {111}},\ \bibinfo {pages}
  {9733--9738} (\bibinfo {year} {2014})} \BibitemShut {NoStop}%
\bibitem [{\citenamefont {Lapin}\ \emph {et~al.}(2006)\citenamefont {Lapin},
  \citenamefont {Schmid},\ and\ \citenamefont {Reuss}}]{Lapin2006}%
  \BibitemOpen
  \bibfield  {author} {\bibinfo {author} {\bibfnamefont {Alexei}\ \bibnamefont
  {Lapin}}, \bibinfo {author} {\bibfnamefont {Joachim}\ \bibnamefont {Schmid}},
  \ and\ \bibinfo {author} {\bibfnamefont {Matthias}\ \bibnamefont {Reuss}},\
  }\bibfield  {title} {\enquote {\bibinfo {title} {{Modeling the dynamics of E.
  coli populations in the three-dimensional turbulent field of a stirred-tank
  bioreactor - A structured-segregated approach}},}\ }\href {\doibase
  10.1016/j.ces.2006.03.003} {\bibfield  {journal} {\bibinfo  {journal}
  {Chemical Engineering Science}\ }\textbf {\bibinfo {volume} {61}},\ \bibinfo
  {pages} {4783--4797} (\bibinfo {year} {2006})}\BibitemShut {NoStop}%
\bibitem [{\citenamefont {Nagylaki}(1980)}]{Nagylaki1980}%
  \BibitemOpen
  \bibfield  {author} {\bibinfo {author} {\bibfnamefont {Thomas}\ \bibnamefont
  {Nagylaki}},\ }\bibfield  {title} {\enquote {\bibinfo {title} {{The
  strong-migration limit in geographically structured populations}},}\ }\href
  {\doibase 10.1007/BF00275916} {\bibfield  {journal} {\bibinfo  {journal}
  {Journal of Mathematical Biology}\ }\textbf {\bibinfo {volume} {9}},\
  \bibinfo {pages} {101--114} (\bibinfo {year} {1980})}\BibitemShut {NoStop}%
\bibitem [{\citenamefont {McPeek}\ and\ \citenamefont
  {Holt}(1992)}]{McPeek1992}%
  \BibitemOpen
  \bibfield  {author} {\bibinfo {author} {\bibfnamefont {Mark~A.}\ \bibnamefont
  {McPeek}}\ and\ \bibinfo {author} {\bibfnamefont {Robert~D.}\ \bibnamefont
  {Holt}},\ }\bibfield  {title} {\enquote {\bibinfo {title} {{The Evolution of
  Dispersal in Spatially and Temporally Varying Environments}},}\ }\href
  {\doibase 10.1086/285453} {\bibfield  {journal} {\bibinfo  {journal} {The
  American Naturalist}\ }\textbf {\bibinfo {volume} {140}},\ \bibinfo {pages}
  {1010--1027} (\bibinfo {year} {1992})}\BibitemShut {NoStop}%
\bibitem [{\citenamefont {Hill}\ \emph {et~al.}(2002)\citenamefont {Hill},
  \citenamefont {Hastings},\ and\ \citenamefont {Botsford}}]{Hill2002a}%
  \BibitemOpen
  \bibfield  {author} {\bibinfo {author} {\bibfnamefont {M.~Forrest}\
  \bibnamefont {Hill}}, \bibinfo {author} {\bibfnamefont {Alan}\ \bibnamefont
  {Hastings}}, \ and\ \bibinfo {author} {\bibfnamefont {Louis~W.}\ \bibnamefont
  {Botsford}},\ }\bibfield  {title} {\enquote {\bibinfo {title} {{The Effects
  of Small Dispersal Rates on Extinction Times in Structured Metapopulation
  Models}},}\ }\href {\doibase 10.1086/341526} {\bibfield  {journal} {\bibinfo
  {journal} {The American Naturalist}\ }\textbf {\bibinfo {volume} {160}},\
  \bibinfo {pages} {389--402} (\bibinfo {year} {2002})}\BibitemShut {NoStop}%
\bibitem [{\citenamefont {Casagrandi}\ and\ \citenamefont
  {Gatto}(2006)}]{Casagrandi2006}%
  \BibitemOpen
  \bibfield  {author} {\bibinfo {author} {\bibfnamefont {Renato}\ \bibnamefont
  {Casagrandi}}\ and\ \bibinfo {author} {\bibfnamefont {Marino}\ \bibnamefont
  {Gatto}},\ }\bibfield  {title} {\enquote {\bibinfo {title} {{The intermediate
  dispersal principle in spatially explicit metapopulations}},}\ }\href
  {\doibase 10.1016/j.jtbi.2005.07.009} {\bibfield  {journal} {\bibinfo
  {journal} {Journal of Theoretical Biology}\ }\textbf {\bibinfo {volume}
  {239}},\ \bibinfo {pages} {22--32} (\bibinfo {year} {2006})}\BibitemShut
  {NoStop}%
\bibitem [{\citenamefont {Houchmandzadeh}\ and\ \citenamefont
  {Vallade}(2011{\natexlab{b}})}]{Houchmandzadeh2011}%
  \BibitemOpen
  \bibfield  {author} {\bibinfo {author} {\bibfnamefont {Bahram}\ \bibnamefont
  {Houchmandzadeh}}\ and\ \bibinfo {author} {\bibfnamefont {Marcel}\
  \bibnamefont {Vallade}},\ }\bibfield  {title} {\enquote {\bibinfo {title}
  {{The fixation probability of a beneficial mutation in a geographically
  structured population}},}\ }\href {\doibase 10.1088/1367-2630/13/7/073020}
  {\bibfield  {journal} {\bibinfo  {journal} {New Journal of Physics}\ }\textbf
  {\bibinfo {volume} {13}},\ \bibinfo {pages} {073020} (\bibinfo {year}
  {2011}{\natexlab{b}})}\BibitemShut {NoStop}%
\bibitem [{\citenamefont {Thalhauser}\ \emph {et~al.}(2010)\citenamefont
  {Thalhauser}, \citenamefont {Lowengrub}, \citenamefont {Stupack},\ and\
  \citenamefont {Komarova}}]{Thalhauser2015}%
  \BibitemOpen
  \bibfield  {author} {\bibinfo {author} {\bibfnamefont {Craig~J.}\
  \bibnamefont {Thalhauser}}, \bibinfo {author} {\bibfnamefont {John~S.}\
  \bibnamefont {Lowengrub}}, \bibinfo {author} {\bibfnamefont {Dwayne}\
  \bibnamefont {Stupack}}, \ and\ \bibinfo {author} {\bibfnamefont
  {Natalia~L.}\ \bibnamefont {Komarova}},\ }\bibfield  {title} {\enquote
  {\bibinfo {title} {{Selection in spatial stochastic models of cancer:
  Migration as a key modulator of fitness}},}\ }\href {\doibase
  10.1186/1745-6150-5-21} {\bibfield  {journal} {\bibinfo  {journal} {Biology
  Direct}\ }\textbf {\bibinfo {volume} {5}},\ \bibinfo {pages} {21} (\bibinfo
  {year} {2010})} \BibitemShut {NoStop}%
\bibitem [{\citenamefont {Krieger}\ \emph {et~al.}(2017)\citenamefont
  {Krieger}, \citenamefont {McAvoy},\ and\ \citenamefont
  {Nowak}}]{Krieger2017}%
  \BibitemOpen
  \bibfield  {author} {\bibinfo {author} {\bibfnamefont {Madison~S.}\
  \bibnamefont {Krieger}}, \bibinfo {author} {\bibfnamefont {Alex}\
  \bibnamefont {McAvoy}}, \ and\ \bibinfo {author} {\bibfnamefont {Martin~A.}\
  \bibnamefont {Nowak}},\ }\bibfield  {title} {\enquote {\bibinfo {title}
  {{Effects of motion in structured populations}},}\ }\href {\doibase
  10.1098/rsif.2017.0509} {\bibfield  {journal} {\bibinfo  {journal} {Journal
  of The Royal Society Interface}\ }\textbf {\bibinfo {volume} {14}},\ \bibinfo
  {pages} {20170509} (\bibinfo {year} {2017})}\BibitemShut {NoStop}%
\bibitem [{\citenamefont {Zimmermann}\ \emph {et~al.}(2004)\citenamefont
  {Zimmermann}, \citenamefont {Egu{\'{i}}luz},\ and\ \citenamefont {{San
  Miguel}}}]{Zimmermann2004}%
  \BibitemOpen
  \bibfield  {author} {\bibinfo {author} {\bibfnamefont {Mart{\'{i}}n~G.}\
  \bibnamefont {Zimmermann}}, \bibinfo {author} {\bibfnamefont
  {V{\'{i}}ctor~M.}\ \bibnamefont {Egu{\'{i}}luz}}, \ and\ \bibinfo {author}
  {\bibfnamefont {Maxi}\ \bibnamefont {{San Miguel}}},\ }\bibfield  {title}
  {\enquote {\bibinfo {title} {{Coevolution of dynamical states and
  interactions in dynamic networks}},}\ }\href {\doibase
  10.1103/PhysRevE.69.065102} {\bibfield  {journal} {\bibinfo  {journal}
  {Physical Review E}\ }\textbf {\bibinfo {volume} {69}},\ \bibinfo {pages}
  {065102} (\bibinfo {year} {2004})}\BibitemShut {NoStop}%
\bibitem [{\citenamefont {Zimmermann}\ and\ \citenamefont
  {Egu{\'{i}}luz}(2005)}]{Zimmermann2005}%
  \BibitemOpen
  \bibfield  {author} {\bibinfo {author} {\bibfnamefont {Mart{\'{i}}n~G.}\
  \bibnamefont {Zimmermann}}\ and\ \bibinfo {author} {\bibfnamefont
  {V{\'{i}}ctor~M.}\ \bibnamefont {Egu{\'{i}}luz}},\ }\bibfield  {title}
  {\enquote {\bibinfo {title} {{Cooperation, social networks, and the emergence
  of leadership in a prisoner's dilemma with adaptive local interactions}},}\
  }\href {\doibase 10.1103/PhysRevE.72.056118} {\bibfield  {journal} {\bibinfo
  {journal} {Physical Review E}\ }\textbf {\bibinfo {volume} {72}},\ \bibinfo
  {pages} {056118} (\bibinfo {year} {2005})}\BibitemShut {NoStop}%
\bibitem [{\citenamefont {Gross}\ \emph {et~al.}(2006)\citenamefont {Gross},
  \citenamefont {D'Lima},\ and\ \citenamefont {Blasius}}]{Gross2006}%
  \BibitemOpen
  \bibfield  {author} {\bibinfo {author} {\bibfnamefont {Thilo}\ \bibnamefont
  {Gross}}, \bibinfo {author} {\bibfnamefont {Carlos J.~Dommar}\ \bibnamefont
  {D'Lima}}, \ and\ \bibinfo {author} {\bibfnamefont {Bernd}\ \bibnamefont
  {Blasius}},\ }\bibfield  {title} {\enquote {\bibinfo {title} {{Epidemic
  Dynamics on an Adaptive Network}},}\ }\href {\doibase
  10.1103/PhysRevLett.96.208701} {\bibfield  {journal} {\bibinfo  {journal}
  {Physical Review Letters}\ }\textbf {\bibinfo {volume} {96}},\ \bibinfo
  {pages} {208701} (\bibinfo {year} {2006})}\BibitemShut {NoStop}%
\bibitem [{\citenamefont {Ehrhardt}\ \emph {et~al.}(2006)\citenamefont
  {Ehrhardt}, \citenamefont {Marsili},\ and\ \citenamefont
  {Vega-Redondo}}]{Ehrhardt2006}%
  \BibitemOpen
  \bibfield  {author} {\bibinfo {author} {\bibfnamefont {George C. M.~A.}\
  \bibnamefont {Ehrhardt}}, \bibinfo {author} {\bibfnamefont {Matteo}\
  \bibnamefont {Marsili}}, \ and\ \bibinfo {author} {\bibfnamefont {Fernando}\
  \bibnamefont {Vega-Redondo}},\ }\bibfield  {title} {\enquote {\bibinfo
  {title} {{Phenomenological models of socioeconomic network dynamics}},}\
  }\href {\doibase 10.1103/PhysRevE.74.036106} {\bibfield  {journal} {\bibinfo
  {journal} {Physical Review E}\ }\textbf {\bibinfo {volume} {74}},\ \bibinfo
  {pages} {036106} (\bibinfo {year} {2006})}\BibitemShut {NoStop}%
\bibitem [{\citenamefont {Pacheco}\ \emph {et~al.}(2006)\citenamefont
  {Pacheco}, \citenamefont {Traulsen},\ and\ \citenamefont
  {Nowak}}]{Pacheco2006}%
  \BibitemOpen
  \bibfield  {author} {\bibinfo {author} {\bibfnamefont {Jorge~M.}\
  \bibnamefont {Pacheco}}, \bibinfo {author} {\bibfnamefont {Arne}\
  \bibnamefont {Traulsen}}, \ and\ \bibinfo {author} {\bibfnamefont
  {Martin~A.}\ \bibnamefont {Nowak}},\ }\bibfield  {title} {\enquote {\bibinfo
  {title} {{Coevolution of Strategy and Structure in Complex Networks with
  Dynamical Linking}},}\ }\href {\doibase 10.1103/PhysRevLett.97.258103}
  {\bibfield  {journal} {\bibinfo  {journal} {Physical Review Letters}\
  }\textbf {\bibinfo {volume} {97}},\ \bibinfo {pages} {258103} (\bibinfo
  {year} {2006})}\BibitemShut {NoStop}%
\bibitem [{\citenamefont {K{\'{a}}rolyi}\ \emph {et~al.}(2005)\citenamefont
  {K{\'{a}}rolyi}, \citenamefont {Neufeld},\ and\ \citenamefont
  {Scheuring}}]{Karolyi2005}%
  \BibitemOpen
  \bibfield  {author} {\bibinfo {author} {\bibfnamefont {Gy{\"{o}}rgy}\
  \bibnamefont {K{\'{a}}rolyi}}, \bibinfo {author} {\bibfnamefont
  {Zolt{\'{a}}n}\ \bibnamefont {Neufeld}}, \ and\ \bibinfo {author}
  {\bibfnamefont {Istv{\'{a}}n}\ \bibnamefont {Scheuring}},\ }\bibfield
  {title} {\enquote {\bibinfo {title} {{Rock-scissors-paper game in a chaotic
  flow: The effect of dispersion on the cyclic competition of
  microorganisms}},}\ }\href {\doibase 10.1016/j.jtbi.2005.02.012} {\bibfield
  {journal} {\bibinfo  {journal} {Journal of Theoretical Biology}\ }\textbf
  {\bibinfo {volume} {236}},\ \bibinfo {pages} {12--20} (\bibinfo {year}
  {2005})}\BibitemShut {NoStop}%
\bibitem [{\citenamefont {Perlekar}\ \emph {et~al.}(2011)\citenamefont
  {Perlekar}, \citenamefont {Benzi}, \citenamefont {Pigolotti},\ and\
  \citenamefont {Toschi}}]{Perlekar2011}%
  \BibitemOpen
  \bibfield  {author} {\bibinfo {author} {\bibfnamefont {Prasad}\ \bibnamefont
  {Perlekar}}, \bibinfo {author} {\bibfnamefont {Roberto}\ \bibnamefont
  {Benzi}}, \bibinfo {author} {\bibfnamefont {Simone}\ \bibnamefont
  {Pigolotti}}, \ and\ \bibinfo {author} {\bibfnamefont {Federico}\
  \bibnamefont {Toschi}},\ }\bibfield  {title} {\enquote {\bibinfo {title}
  {{Particle algorithms for population dynamics in flows}},}\ }\href {\doibase
  10.1088/1742-6596/333/1/012013} {\bibfield  {journal} {\bibinfo  {journal}
  {Journal of Physics: Conference Series}\ }\textbf {\bibinfo {volume} {333}},\
  \bibinfo {pages} {012013} (\bibinfo {year} {2011})}\BibitemShut {NoStop}%
\bibitem [{\citenamefont {Pigolotti}\ \emph {et~al.}(2012)\citenamefont
  {Pigolotti}, \citenamefont {Benzi}, \citenamefont {Jensen},\ and\
  \citenamefont {Nelson}}]{Pigolotti2012}%
  \BibitemOpen
  \bibfield  {author} {\bibinfo {author} {\bibfnamefont {Simone}\ \bibnamefont
  {Pigolotti}}, \bibinfo {author} {\bibfnamefont {Roberto}\ \bibnamefont
  {Benzi}}, \bibinfo {author} {\bibfnamefont {Mogens~H.}\ \bibnamefont
  {Jensen}}, \ and\ \bibinfo {author} {\bibfnamefont {David~R.}\ \bibnamefont
  {Nelson}},\ }\bibfield  {title} {\enquote {\bibinfo {title} {{Population
  Genetics in Compressible Flows}},}\ }\href {\doibase
  10.1103/PhysRevLett.108.128102} {\bibfield  {journal} {\bibinfo  {journal}
  {Physical Review Letters}\ }\textbf {\bibinfo {volume} {108}},\ \bibinfo
  {pages} {128102} (\bibinfo {year} {2012})}\BibitemShut {NoStop}%
\bibitem [{\citenamefont {Benzi}\ \emph {et~al.}(2012)\citenamefont {Benzi},
  \citenamefont {Jensen}, \citenamefont {Nelson}, \citenamefont {Perlekar},
  \citenamefont {Pigolotti},\ and\ \citenamefont {Toschi}}]{Pigolotti2012a}%
  \BibitemOpen
  \bibfield  {author} {\bibinfo {author} {\bibfnamefont {Roberto}\ \bibnamefont
  {Benzi}}, \bibinfo {author} {\bibfnamefont {Mogens~H.}\ \bibnamefont
  {Jensen}}, \bibinfo {author} {\bibfnamefont {David~R.}\ \bibnamefont
  {Nelson}}, \bibinfo {author} {\bibfnamefont {Prasad}\ \bibnamefont
  {Perlekar}}, \bibinfo {author} {\bibfnamefont {Simone}\ \bibnamefont
  {Pigolotti}}, \ and\ \bibinfo {author} {\bibfnamefont {Federico}\
  \bibnamefont {Toschi}},\ }\bibfield  {title} {\enquote {\bibinfo {title}
  {{Population dynamics in compressible flows}},}\ }\href {\doibase
  10.1140/epjst/e2012-01552-0} {\bibfield  {journal} {\bibinfo  {journal} {The
  European Physical Journal Special Topics}\ }\textbf {\bibinfo {volume}
  {204}},\ \bibinfo {pages} {57--73} (\bibinfo {year} {2012})}\BibitemShut
  {NoStop}%
\bibitem [{\citenamefont {Pigolotti}\ \emph {et~al.}(2013)\citenamefont
  {Pigolotti}, \citenamefont {Benzi}, \citenamefont {Perlekar}, \citenamefont
  {Jensen}, \citenamefont {Toschi},\ and\ \citenamefont
  {Nelson}}]{Pigolotti2013}%
  \BibitemOpen
  \bibfield  {author} {\bibinfo {author} {\bibfnamefont {Simone}\ \bibnamefont
  {Pigolotti}}, \bibinfo {author} {\bibfnamefont {Roberto}\ \bibnamefont
  {Benzi}}, \bibinfo {author} {\bibfnamefont {Prasad}\ \bibnamefont
  {Perlekar}}, \bibinfo {author} {\bibfnamefont {Mogens~H.}\ \bibnamefont
  {Jensen}}, \bibinfo {author} {\bibfnamefont {Federico}\ \bibnamefont
  {Toschi}}, \ and\ \bibinfo {author} {\bibfnamefont {David~R.}\ \bibnamefont
  {Nelson}},\ }\bibfield  {title} {\enquote {\bibinfo {title} {{Growth,
  competition and cooperation in spatial population genetics}},}\ }\href
  {\doibase 10.1016/j.tpb.2012.12.002} {\bibfield  {journal} {\bibinfo
  {journal} {Theoretical Population Biology}\ }\textbf {\bibinfo {volume}
  {84}},\ \bibinfo {pages} {72--86} (\bibinfo {year} {2013})}\BibitemShut
  {NoStop}%
\bibitem [{\citenamefont {Pigolotti}\ and\ \citenamefont
  {Benzi}(2014)}]{Pigolotti2014}%
  \BibitemOpen
  \bibfield  {author} {\bibinfo {author} {\bibfnamefont {Simone}\ \bibnamefont
  {Pigolotti}}\ and\ \bibinfo {author} {\bibfnamefont {Roberto}\ \bibnamefont
  {Benzi}},\ }\bibfield  {title} {\enquote {\bibinfo {title} {{Selective
  Advantage of Diffusing Faster}},}\ }\href {\doibase
  10.1103/PhysRevLett.112.188102} {\bibfield  {journal} {\bibinfo  {journal}
  {Physical Review Letters}\ }\textbf {\bibinfo {volume} {112}},\ \bibinfo
  {pages} {188102} (\bibinfo {year} {2014})}\BibitemShut {NoStop}%
\bibitem [{\citenamefont {Plummer}\ \emph {et~al.}(2018)\citenamefont
  {Plummer}, \citenamefont {Benzi}, \citenamefont {Nelson},\ and\ \citenamefont
  {Toschi}}]{Plummer2018}%
  \BibitemOpen
  \bibfield  {author} {\bibinfo {author} {\bibfnamefont {Abigail}\ \bibnamefont
  {Plummer}}, \bibinfo {author} {\bibfnamefont {Roberto}\ \bibnamefont
  {Benzi}}, \bibinfo {author} {\bibfnamefont {David~R.}\ \bibnamefont
  {Nelson}}, \ and\ \bibinfo {author} {\bibfnamefont {Federico}\ \bibnamefont
  {Toschi}},\ }\bibfield  {title} {\enquote {\bibinfo {title} {{Fixation
  probabilities in weakly compressible fluid flows}},}\ }\href
  {\doibase 10.1073/pnas.1812829116} {\bibfield  {journal} {\bibinfo  {journal} 
  {Proceedings of the National Academy of Sciences}\  }\textbf {\bibinfo {volume}
  {116}},\ \bibinfo {pages} {373--378} (\bibinfo  {year} {2019})}\BibitemShut {NoStop}%
\bibitem [{\citenamefont {Minors}\ \emph {et~al.}(2018)\citenamefont {Minors},
  \citenamefont {Rogers},\ and\ \citenamefont {Yates}}]{Minors2018}%
  \BibitemOpen
  \bibfield  {author} {\bibinfo {author} {\bibfnamefont {Kevin}\ \bibnamefont
  {Minors}}, \bibinfo {author} {\bibfnamefont {Tim}\ \bibnamefont {Rogers}}, \
  and\ \bibinfo {author} {\bibfnamefont {Christian~A.}\ \bibnamefont {Yates}},\
  }\bibfield  {title} {\enquote {\bibinfo {title} {{Noise-driven bias in the
  non-local voter model}},}\ }\href {\doibase 10.1209/0295-5075/122/10004}
  {\bibfield  {journal} {\bibinfo  {journal} {EPL (Europhysics Letters)}\
  }\textbf {\bibinfo {volume} {122}},\ \bibinfo {pages} {10004} (\bibinfo
  {year} {2018})}\BibitemShut {NoStop}%
\bibitem [{\citenamefont {Herrer{\'{i}}as-Azcu{\'{e}}}\ \emph
  {et~al.}(2018)\citenamefont {Herrer{\'{i}}as-Azcu{\'{e}}}, \citenamefont
  {P{\'{e}}rez-Mu{\~{n}}uzuri},\ and\ \citenamefont {Galla}}]{Herrerias2018}%
  \BibitemOpen
  \bibfield  {author} {\bibinfo {author} {\bibfnamefont {Francisco}\
  \bibnamefont {Herrer{\'{i}}as-Azcu{\'{e}}}}, \bibinfo {author} {\bibfnamefont
  {Vicente}\ \bibnamefont {P{\'{e}}rez-Mu{\~{n}}uzuri}}, \ and\ \bibinfo
  {author} {\bibfnamefont {Tobias}\ \bibnamefont {Galla}},\ }\bibfield  {title}
  {\enquote {\bibinfo {title} {{Stirring does not make populations well
  mixed}},}\ }\href {\doibase 10.1038/s41598-018-22062-w} {\bibfield  {journal}
  {\bibinfo  {journal} {Scientific Reports}\ }\textbf {\bibinfo {volume} {8}},\
  \bibinfo {pages} {4068} (\bibinfo {year} {2018})}\BibitemShut {NoStop}%
\bibitem [{\citenamefont {Ottino}(1989)}]{Ottino1989}%
  \BibitemOpen
  \bibfield  {author} {\bibinfo {author} {\bibfnamefont {Julio~M.}\
  \bibnamefont {Ottino}},\ }\href@noop {} {\emph {\bibinfo {title} {{The
  kinematics of mixing: stretching, chaos, and transport}}}}\ (\bibinfo
  {publisher} {Cambridge University Press},\ \bibinfo {address} {Cambridge,
  UK},\ \bibinfo {year} {1989})\BibitemShut {NoStop}%
\bibitem [{\citenamefont {Neufeld}\ and\ \citenamefont
  {Hern{\'{a}}ndez-Garc{\'{i}}a}(2010)}]{Neufeld2010}%
  \BibitemOpen
  \bibfield  {author} {\bibinfo {author} {\bibfnamefont {Zolt{\'{a}}n}\
  \bibnamefont {Neufeld}}\ and\ \bibinfo {author} {\bibfnamefont {Emilio}\
  \bibnamefont {Hern{\'{a}}ndez-Garc{\'{i}}a}},\ }\href@noop {} {\emph
  {\bibinfo {title} {{Chemical and biological processes in fluid flows: a
  dynamical systems approach}}}}\ (\bibinfo  {publisher} {Imperial College
  Press},\ \bibinfo {address} {London, UK},\ \bibinfo {year}
  {2010})\BibitemShut {NoStop}%
\bibitem [{\citenamefont {P{\'{e}}rez-Mu{\~{n}}uzuri}\ and\ \citenamefont
  {Fern{\'{a}}ndez-Garc{\'{i}}a}(2007)}]{Perez-Munuzuri2007}%
  \BibitemOpen
  \bibfield  {author} {\bibinfo {author} {\bibfnamefont {Vicente}\ \bibnamefont
  {P{\'{e}}rez-Mu{\~{n}}uzuri}}\ and\ \bibinfo {author} {\bibfnamefont
  {Guillermo}\ \bibnamefont {Fern{\'{a}}ndez-Garc{\'{i}}a}},\ }\bibfield
  {title} {\enquote {\bibinfo {title} {{Mixing efficiency in an excitable
  medium with chaotic shear flow}},}\ }\href {\doibase
  10.1103/PhysRevE.75.046209} {\bibfield  {journal} {\bibinfo  {journal}
  {Physical Review E}\ }\textbf {\bibinfo {volume} {75}},\ \bibinfo {pages}
  {046209} (\bibinfo {year} {2007})}\BibitemShut {NoStop}%
\bibitem [{\citenamefont {Galla}\ and\ \citenamefont
  {P{\'{e}}rez-Mu{\~{n}}uzuri}(2017)}]{Galla2016}%
  \BibitemOpen
  \bibfield  {author} {\bibinfo {author} {\bibfnamefont {Tobias}\ \bibnamefont
  {Galla}}\ and\ \bibinfo {author} {\bibfnamefont {Vicente}\ \bibnamefont
  {P{\'{e}}rez-Mu{\~{n}}uzuri}},\ }\bibfield  {title} {\enquote {\bibinfo
  {title} {{Time scales and species coexistence in chaotic flows}},}\ }\href
  {\doibase 10.1209/0295-5075/117/68001} {\bibfield  {journal} {\bibinfo
  {journal} {EPL (Europhysics Letters)}\ }\textbf {\bibinfo {volume} {117}},\
  \bibinfo {pages} {68001} (\bibinfo {year} {2017})}\BibitemShut {NoStop}%
\bibitem [{\citenamefont {Young}\ \emph {et~al.}(2001)\citenamefont {Young},
  \citenamefont {Roberts},\ and\ \citenamefont {Stuhne}}]{Young2001}%
  \BibitemOpen
  \bibfield  {author} {\bibinfo {author} {\bibfnamefont {William~R.}\
  \bibnamefont {Young}}, \bibinfo {author} {\bibfnamefont {Anthony~J.}\
  \bibnamefont {Roberts}}, \ and\ \bibinfo {author} {\bibfnamefont {Gordan~R.}\
  \bibnamefont {Stuhne}},\ }\bibfield  {title} {\enquote {\bibinfo {title}
  {{Reproductive pair correlations and the clustering of organisms}},}\ }\href
  {\doibase 10.1038/35085561} {\bibfield  {journal} {\bibinfo  {journal}
  {Nature}\ }\textbf {\bibinfo {volume} {412}},\ \bibinfo {pages} {328--331}
  (\bibinfo {year} {2001})}\BibitemShut {NoStop}%
\bibitem [{\citenamefont {Sandulescu}\ \emph {et~al.}(2007)\citenamefont
  {Sandulescu}, \citenamefont {L{\'{o}}pez}, \citenamefont
  {Hern{\'{a}}ndez-Garc{\'{i}}a},\ and\ \citenamefont
  {Feudel}}]{Sandulescu2007}%
  \BibitemOpen
  \bibfield  {author} {\bibinfo {author} {\bibfnamefont {Mathias}\ \bibnamefont
  {Sandulescu}}, \bibinfo {author} {\bibfnamefont {Crist{\'{o}}bal}\
  \bibnamefont {L{\'{o}}pez}}, \bibinfo {author} {\bibfnamefont {Emilio}\
  \bibnamefont {Hern{\'{a}}ndez-Garc{\'{i}}a}}, \ and\ \bibinfo {author}
  {\bibfnamefont {Ulrike}\ \bibnamefont {Feudel}},\ }\bibfield  {title}
  {\enquote {\bibinfo {title} {{Plankton blooms in vortices: the role of
  biological and hydrodynamic timescales}},}\ }\href {\doibase
  10.5194/npg-14-443-2007} {\bibfield  {journal} {\bibinfo  {journal}
  {Nonlinear Processes in Geophysics}\ }\textbf {\bibinfo {volume} {14}},\
  \bibinfo {pages} {443--454} (\bibinfo {year} {2007})}\BibitemShut {NoStop}%
\bibitem [{\citenamefont {Neufeld}\ \emph {et~al.}(2002)\citenamefont
  {Neufeld}, \citenamefont {Haynes},\ and\ \citenamefont
  {T{\'{e}}l}}]{Neufeld2002}%
  \BibitemOpen
  \bibfield  {author} {\bibinfo {author} {\bibfnamefont {Zolt{\'{a}}n}\
  \bibnamefont {Neufeld}}, \bibinfo {author} {\bibfnamefont {Peter~H.}\
  \bibnamefont {Haynes}}, \ and\ \bibinfo {author} {\bibfnamefont
  {Tam{\'{a}}s}\ \bibnamefont {T{\'{e}}l}},\ }\bibfield  {title} {\enquote
  {\bibinfo {title} {{Chaotic mixing induced transitions in
  reaction–diffusion systems}},}\ }\href {\doibase 10.1063/1.1476949}
  {\bibfield  {journal} {\bibinfo  {journal} {Chaos: An Interdisciplinary
  Journal of Nonlinear Science}\ }\textbf {\bibinfo {volume} {12}},\ \bibinfo
  {pages} {426--438} (\bibinfo {year} {2002})}\BibitemShut {NoStop}%
\bibitem [{\citenamefont {Gilbert}(1961)}]{Gilbert1961}%
  \BibitemOpen
  \bibfield  {author} {\bibinfo {author} {\bibfnamefont {Edgar~N.}\
  \bibnamefont {Gilbert}},\ }\bibfield  {title} {\enquote {\bibinfo {title}
  {{Random Plane Networks}},}\ }\href {\doibase 10.1137/0109045} {\bibfield
  {journal} {\bibinfo  {journal} {Journal of the Society for Industrial and
  Applied Mathematics}\ }\textbf {\bibinfo {volume} {9}},\ \bibinfo {pages}
  {533--543} (\bibinfo {year} {1961})}\BibitemShut {NoStop}%
\bibitem [{\citenamefont {Hindersin}\ \emph {et~al.}(2016)\citenamefont
  {Hindersin}, \citenamefont {M{\"{o}}ller}, \citenamefont {Traulsen},\ and\
  \citenamefont {Bauer}}]{Hindersin2016}%
  \BibitemOpen
  \bibfield  {author} {\bibinfo {author} {\bibfnamefont {Laura}\ \bibnamefont
  {Hindersin}}, \bibinfo {author} {\bibfnamefont {Marius}\ \bibnamefont
  {M{\"{o}}ller}}, \bibinfo {author} {\bibfnamefont {Arne}\ \bibnamefont
  {Traulsen}}, \ and\ \bibinfo {author} {\bibfnamefont {Benedikt}\ \bibnamefont
  {Bauer}},\ }\bibfield  {title} {\enquote {\bibinfo {title} {{Exact numerical
  calculation of fixation probability and time on graphs}},}\ }\href {\doibase
  10.1016/j.biosystems.2016.08.010} {\bibfield  {journal} {\bibinfo  {journal}
  {Biosystems}\ }\textbf {\bibinfo {volume} {150}},\ \bibinfo {pages} {87--91}
  (\bibinfo {year} {2016})} \BibitemShut {NoStop}%
\bibitem [{\citenamefont {Giakkoupis}(2016)}]{Giakkoupis2016}%
  \BibitemOpen
  \bibfield  {author} {\bibinfo {author} {\bibfnamefont {George}\ \bibnamefont
  {Giakkoupis}},\ }\bibfield  {title} {\enquote {\bibinfo {title} {{Amplifiers
  and Suppressors of Selection for the Moran Process on Undirected Graphs}},}\
  }\href {http://arxiv.org/abs/1611.01585} {\bibfield  {journal} {\bibinfo
  {journal} {arXiv preprint}\ ,\ \bibinfo {pages} {1--27}} (\bibinfo {year}
  {2016})}
  \BibitemShut {NoStop}%
\bibitem [{\citenamefont {Adlam}\ \emph {et~al.}(2015)\citenamefont {Adlam},
  \citenamefont {Chatterjee},\ and\ \citenamefont {Nowak}}]{Adlam2015}%
  \BibitemOpen
  \bibfield  {author} {\bibinfo {author} {\bibfnamefont {Ben}\ \bibnamefont
  {Adlam}}, \bibinfo {author} {\bibfnamefont {Krishnendu}\ \bibnamefont
  {Chatterjee}}, \ and\ \bibinfo {author} {\bibfnamefont {Martin~A.}\
  \bibnamefont {Nowak}},\ }\bibfield  {title} {\enquote {\bibinfo {title}
  {{Amplifiers of selection}},}\ }\href {\doibase 10.1098/rspa.2015.0114}
  {\bibfield  {journal} {\bibinfo  {journal} {Proceedings of the Royal Society
  A: Mathematical, Physical and Engineering Science}\ }\textbf {\bibinfo
  {volume} {471}},\ \bibinfo {pages} {20150114} (\bibinfo {year}
  {2015})}\BibitemShut {NoStop}%
\bibitem [{\citenamefont {Antal}\ \emph {et~al.}(2006)\citenamefont {Antal},
  \citenamefont {Redner},\ and\ \citenamefont {Sood}}]{Antal2006}%
  \BibitemOpen
  \bibfield  {author} {\bibinfo {author} {\bibfnamefont {Tibor}\ \bibnamefont
  {Antal}}, \bibinfo {author} {\bibfnamefont {Sidney}\ \bibnamefont {Redner}},
  \ and\ \bibinfo {author} {\bibfnamefont {Vishal}\ \bibnamefont {Sood}},\
  }\bibfield  {title} {\enquote {\bibinfo {title} {{Evolutionary Dynamics on
  Degree-Heterogeneous Graphs}},}\ }\href {\doibase
  10.1103/PhysRevLett.96.188104} {\bibfield  {journal} {\bibinfo  {journal}
  {Physical Review Letters}\ }\textbf {\bibinfo {volume} {96}},\ \bibinfo
  {pages} {188104} (\bibinfo {year} {2006})}\BibitemShut {NoStop}%
\bibitem [{\citenamefont {Sood}\ \emph {et~al.}(2008)\citenamefont {Sood},
  \citenamefont {Antal},\ and\ \citenamefont {Redner}}]{Sood2008}%
  \BibitemOpen
  \bibfield  {author} {\bibinfo {author} {\bibfnamefont {Vishal}\ \bibnamefont
  {Sood}}, \bibinfo {author} {\bibfnamefont {Tibor}\ \bibnamefont {Antal}}, \
  and\ \bibinfo {author} {\bibfnamefont {Sidney}\ \bibnamefont {Redner}},\
  }\bibfield  {title} {\enquote {\bibinfo {title} {{Voter models on
  heterogeneous networks}},}\ }\href {\doibase 10.1103/PhysRevE.77.041121}
  {\bibfield  {journal} {\bibinfo  {journal} {Physical Review E}\ }\textbf
  {\bibinfo {volume} {77}},\ \bibinfo {pages} {041121} (\bibinfo {year}
  {2008})}\BibitemShut {NoStop}%
\bibitem [{\citenamefont {Maciejewski}(2014)}]{Maciejewski2014}%
  \BibitemOpen
  \bibfield  {author} {\bibinfo {author} {\bibfnamefont {Wes}\ \bibnamefont
  {Maciejewski}},\ }\bibfield  {title} {\enquote {\bibinfo {title}
  {{Reproductive value in graph-structured populations}},}\ }\href {\doibase
  10.1016/j.jtbi.2013.09.032} {\bibfield  {journal} {\bibinfo  {journal}
  {Journal of Theoretical Biology}\ }\textbf {\bibinfo {volume} {340}},\
  \bibinfo {pages} {285--293} (\bibinfo {year} {2014})}\BibitemShut {NoStop}%
\bibitem [{\citenamefont {Tan}\ and\ \citenamefont {L{\"{u}}}(2015)}]{Tan2014}%
  \BibitemOpen
  \bibfield  {author} {\bibinfo {author} {\bibfnamefont {Shaolin}\ \bibnamefont
  {Tan}}\ and\ \bibinfo {author} {\bibfnamefont {Jinhu}\ \bibnamefont
  {L{\"{u}}}},\ }\bibfield  {title} {\enquote {\bibinfo {title}
  {{Characterizing the effect of population heterogeneity on evolutionary
  dynamics on complex networks}},}\ }\href {\doibase 10.1038/srep05034}
  {\bibfield  {journal} {\bibinfo  {journal} {Scientific Reports}\ }\textbf
  {\bibinfo {volume} {4}},\ \bibinfo {pages} {5034} (\bibinfo {year}
  {2015})}\BibitemShut {NoStop}%
\bibitem [{\citenamefont {Ma}\ and\ \citenamefont {Hanna}(1999)}]{Ma1999}%
  \BibitemOpen
  \bibfield  {author} {\bibinfo {author} {\bibfnamefont {Fangrui}\ \bibnamefont
  {Ma}}\ and\ \bibinfo {author} {\bibfnamefont {Milford~A.}\ \bibnamefont
  {Hanna}},\ }\bibfield  {title} {\enquote {\bibinfo {title} {{Biodiesel
  production: a review}},}\ }\href {\doibase 10.1016/S0960-8524(99)00025-5}
  {\bibfield  {journal} {\bibinfo  {journal} {Bioresource Technology}\ }\textbf
  {\bibinfo {volume} {70}},\ \bibinfo {pages} {1--15} (\bibinfo {year}
  {1999})}\BibitemShut {NoStop}%
\bibitem [{\citenamefont {Chisti}(2007)}]{Chisti2007a}%
  \BibitemOpen
  \bibfield  {author} {\bibinfo {author} {\bibfnamefont {Yusuf}\ \bibnamefont
  {Chisti}},\ }\bibfield  {title} {\enquote {\bibinfo {title} {{Biodiesel from
  microalgae}},}\ }\href {\doibase 10.1016/j.biotechadv.2007.02.001} {\bibfield
   {journal} {\bibinfo  {journal} {Biotechnology Advances}\ }\textbf {\bibinfo
  {volume} {25}},\ \bibinfo {pages} {294--306} (\bibinfo {year} {2007})}
  \BibitemShut {NoStop}%
\bibitem [{\citenamefont {Jeon}\ and\ \citenamefont {Yeom}(2010)}]{Jeon2010a}%
  \BibitemOpen
  \bibfield  {author} {\bibinfo {author} {\bibfnamefont {Dong~Jin}\
  \bibnamefont {Jeon}}\ and\ \bibinfo {author} {\bibfnamefont {Sung~Ho}\
  \bibnamefont {Yeom}},\ }\bibfield  {title} {\enquote {\bibinfo {title}
  {{Two-step bioprocess employing whole cell and enzyme for economical
  biodiesel production}},}\ }\href {\doibase 10.1007/s11814-010-0263-y}
  {\bibfield  {journal} {\bibinfo  {journal} {Korean Journal of Chemical
  Engineering}\ }\textbf {\bibinfo {volume} {27}},\ \bibinfo {pages}
  {1555--1559} (\bibinfo {year} {2010})}\BibitemShut {NoStop}%
\bibitem [{\citenamefont {Meng}\ \emph {et~al.}(2009)\citenamefont {Meng},
  \citenamefont {Yang}, \citenamefont {Xu}, \citenamefont {Zhang},
  \citenamefont {Nie},\ and\ \citenamefont {Xian}}]{Meng2009a}%
  \BibitemOpen
  \bibfield  {author} {\bibinfo {author} {\bibfnamefont {Xin}\ \bibnamefont
  {Meng}}, \bibinfo {author} {\bibfnamefont {Jianming}\ \bibnamefont {Yang}},
  \bibinfo {author} {\bibfnamefont {Xin}\ \bibnamefont {Xu}}, \bibinfo {author}
  {\bibfnamefont {Lei}\ \bibnamefont {Zhang}}, \bibinfo {author} {\bibfnamefont
  {Qingjuan}\ \bibnamefont {Nie}}, \ and\ \bibinfo {author} {\bibfnamefont
  {Mo}~\bibnamefont {Xian}},\ }\bibfield  {title} {\enquote {\bibinfo {title}
  {{Biodiesel production from oleaginous microorganisms}},}\ }\href {\doibase
  10.1016/j.renene.2008.04.014} {\bibfield  {journal} {\bibinfo  {journal}
  {Renewable Energy}\ }\textbf {\bibinfo {volume} {34}},\ \bibinfo {pages}
  {1--5} (\bibinfo {year} {2009})}
  \BibitemShut {NoStop}%
\bibitem [{\citenamefont {Wahl}\ \emph {et~al.}(2002)\citenamefont {Wahl},
  \citenamefont {Gerrish},\ and\ \citenamefont {Saika-Voivod}}]{Wahl2002}%
  \BibitemOpen
  \bibfield  {author} {\bibinfo {author} {\bibfnamefont {Lindi~M.}\
  \bibnamefont {Wahl}}, \bibinfo {author} {\bibfnamefont {Philip~J}\
  \bibnamefont {Gerrish}}, \ and\ \bibinfo {author} {\bibfnamefont {Ivan}\
  \bibnamefont {Saika-Voivod}},\ }\bibfield  {title} {\enquote {\bibinfo
  {title} {{Evaluating the impact of population bottlenecks in experimental
  evolution.}}}\ }\href {http://www.ncbi.nlm.nih.gov/pubmed/12399403
  http://www.pubmedcentral.nih.gov/articlerender.fcgi?artid=PMC1462272}
  {\bibfield  {journal} {\bibinfo  {journal} {Genetics}\ }\textbf {\bibinfo
  {volume} {162}},\ \bibinfo {pages} {961--71} (\bibinfo {year}
  {2002})}\BibitemShut {NoStop}%
\bibitem [{\citenamefont {Patwa}\ and\ \citenamefont {Wahl}(2008)}]{Patwa2008}%
  \BibitemOpen
  \bibfield  {author} {\bibinfo {author} {\bibfnamefont {Zaheerabbas}\
  \bibnamefont {Patwa}}\ and\ \bibinfo {author} {\bibfnamefont {Linda~M.}\
  \bibnamefont {Wahl}},\ }\bibfield  {title} {\enquote {\bibinfo {title} {{The
  fixation probability of beneficial mutations}},}\ }\href {\doibase
  10.1098/rsif.2008.0248} {\bibfield  {journal} {\bibinfo  {journal} {Journal
  of The Royal Society Interface}\ }\textbf {\bibinfo {volume} {5}},\ \bibinfo
  {pages} {1279--1289} (\bibinfo {year} {2008})}\BibitemShut {NoStop}%
\end{thebibliography}
\end{document}